\documentclass[pre,showpacs]{revtex4}
\usepackage{epsf}
\usepackage{amssymb}
\begin{document}
\newcommand {\vtd}[2]   {{\lh\!\bay{c}#1\\#2\eay\!\rh}}
\newcommand {\nl  } {\newline        } \newcommand {\nn  } {\nonumber     }
\newcommand {\noi } {\noindent       } \newcommand {\np  } {\newpage      }
\newcommand {\ov  } {\over        }    \newcommand {\pa  } {\partial      }
\newcommand {\e   } {\!+\!           } \newcommand {\m   } {\!-\!         }
\newcommand {\Bra } {\left\langle    } \newcommand {\Ket } {\right\rangle }
\newcommand {\bra } {     \langle    } \newcommand {\ket } {      \rangle }
\newcommand {\beq } {\begin{equation}} \newcommand {\eeq } {\end{equation}}
\newcommand {\bea } {\begin{eqnarray}} \newcommand {\eea } {\end{eqnarray}}
\newcommand {\bay } {\begin{array}   } \newcommand {\eay } {\end{array}   }
\newcommand {\lh  } {\left(          } \newcommand {\rh  } {\right)       }
\newcommand {\lv  } {\left[          } \newcommand {\rv  } {\right]       }
\newcommand {\lc  } {\left\{         } \newcommand {\rc  } {\right\}      }
\newcommand {\lp  } {\left.          } \newcommand {\rp  } {\right.       }
\newcommand {\de  } {\delta          } \newcommand {\De  } {\Delta        } 
\newcommand {\om  } {\omega          } \newcommand {\be  } {\beta         }
\newcommand {\rmm } {{\rm    m}      } \newcommand {\tin } {{\tilde{\rm n}}}
\newcommand {\ri  } {{\rm    i}      } \newcommand {\rn  } {{\rm    n}    }
\newcommand {\htom} {{\hat{\om }}}     \newcommand {\htpi} {{\hat{\pi }}}
\newcommand {\htrh} {{\hat{\rho}}}     \newcommand {\hty } {{\hat{y   }}}
\newcommand {\htm } {{\hat{m   }}}     \newcommand {\htq } {{\hat{q   }}}
\newcommand {\vm  } {{\vec{m   }}}     \newcommand {\vM  } {{\vec{M   }}}
\newcommand {\vx }  {{\vec{x   }}}     \newcommand {\htx } {{\hat{x   }}}
                                       \newcommand {\vhx } {{\vec{\hat{x}}}} 
\newcommand {\vv  } {{\vec{v   }}}     \newcommand {\htv } {{\hat{v   }}}
                                       \newcommand {\vhv } {{\vec{\hat{v}}}}    
\newcommand {\vf  } {{\vec{f   }}}     \newcommand {\htf } {{\hat{f   }}}     
\newcommand {\vhf } {{\vec{\hat{f}}}}
\newcommand {\htc } {{\hat{c   }}}     \newcommand {\htd } {{\hat{d   }}}
\newcommand {\vcc } {{\vec{c   }}}     \newcommand {\vn  } {{\vec{n   }}}
\newcommand {\vcr } {{\vec{r   }}}     \newcommand {\vs  } {{\vec{s   }}}
\newcommand {\vt  } {{\vec{t   }}}     \newcommand {\vz  } {{\vec{z   }}}
\newcommand {\vy  } {{\vec{y   }}}
\newcommand {\cC  } {{\cal   C }}      \newcommand {\cE  } {{\cal   E }}       
\newcommand {\cD  } {{\cal   D }}      \newcommand {\cF  } {{\cal   F }}       
\newcommand {\cG  } {{\cal   G }}      \newcommand {\cI  } {{\cal   I }}
\newcommand {\cH  } {{\cal   H }}      
\newcommand {\cM  } {{\cal   M }}      \newcommand {\cN  } {{\cal   N }}       
\newcommand {\cO  } {{\cal   O }}      \newcommand {\cQ  } {{\cal   Q }}       
\newcommand {\cS  } {{\cal   S }}      \newcommand {\cT  } {{\cal   T }}      
\newcommand {\cU  } {{\cal   U }}      \newcommand {\cW  } {{\cal   W }}      
\newcommand {\cX  } {{\cal   X }}      \newcommand {\cZ  } {{\cal   Z }}      
\newcommand {\bfA } {{\bf    A }}      \newcommand {\bfB } {{\bf    B }}
\newcommand {\bfC } {{\bf    C }}      \newcommand {\bfG } {{\bf    G }}
\newcommand {\BKK } {{\Bra   \Ket_K}}  \newcommand {\BKm } {{\Bra a \Ket_\rmm}}
                                       \newcommand {\BKmu} {{\{ \mu \}_\rmm}}
\newcommand {\eps } {{\varepsilon}}
\newcommand {\la  } {{\lambda}}
\newcommand {\Tr  } {\mathop{\mbox{\rm Tr   }}}
\newcommand {\ext } {\mathop{\mbox{\rm ext  }}}
\newcommand {\ath } {\mathop{\mbox{\rm atanh}}}
\newcommand {\fns } {\footnotesize}
\newcommand {\hsc } {\hspace*{1cm} }
\newcommand {\ha  } {{1\over 2}}       \newcommand {\ev  } {\equiv        }
\newcommand {\BK  } {{\Bra\Ket}}
\newcommand {\oL  } {{\overline{L}}} 
\newcommand {\cP  } {{\cal{P}}} 
\newcommand {\no  } {\neg} 
\newcommand {\ftd}[2]   {\fns{\lh\!\bay{c}#1\\#2\eay\!\rh}}
\newsavebox{\uuunit}
\sbox{\uuunit}
    {\setlength{\unitlength}{0.825em}
     \begin{picture}(0.6,0.7)
        \thinlines
        \put(0,0){\line(1,0){0.5}}
        \put(0.15,0){\line(0,1){0.7}}
        \put(0.35,0){\line(0,1){0.8}}
       \multiput(0.3,0.8)(-0.04,-0.02){12}{\rule{0.5pt}{0.5pt}}
     \end {picture}}
\newcommand {\Unity}{\mathord{\!\usebox{\uuunit}}}

\title{Random Graph Coloring - a Statistical Physics Approach}

\author{J. van~Mourik and D.~Saad}
\affiliation{
The Neural Computing Research Group,  Aston University,
Birmingham B4 7ET, United Kingdom}

\begin{abstract}
The problem of vertex coloring in random graphs is studied using
methods of statistical physics and probability.  Our analytical
results are compared to those obtained by exact enumeration and
Monte-Carlo simulations. We critically discuss the merits and
shortcomings of the various methods, and interpret the results
obtained. We present an exact analytical expression for the 2-coloring
problem as well as general replica symmetric approximated solutions
for the thermodynamics of the graph coloring problem with $p$ colors
and $K$-body edges. 
\end{abstract}
%
\maketitle
%
\section{Introduction \label{sec:IN}}
%
Methods of statistical physics have recently been applied to a variety of 
complex optimization problems in a broad range of areas, from computational 
complexity~\cite{SAT,vertexcovering} to the study of error correcting 
codes~\cite{errc} and cryptography~\cite{us_prl,hidingstates_prl}.

Graph coloring is one of the basic Non-deterministically Polynomial (NP) 
problems. The task is to assign one of $p$-colors to each node, in a randomly 
connected set of vertices, such that no edge will have the same colors 
assigned to both ends. The feasibility of finding such a solution clearly 
depends on the level and nature of connectivity in the graph and the number of 
colors. The very existence of a solution is in the class of NP-complete 
problems~\cite{GareyJohnson}. An extension of the problem to the case of 
hyper-edges comprising more than two vertices is also of practical 
significance~\cite{graphcoloring}.

Recent success in the application of statistical physics techniques to
computational complexity problems, naturally lead to the belief that 
they may be applied to a wide range of computational complexity tasks,
among them is the graph coloring problem. 

In this paper we map the graph coloring problem, with $p$-colors, onto
the anti-ferromagnetic $p$-spin Potts model~\cite{antiferroPotts}; this 
facilitates the use of established methods of statistical physics for gaining 
insight into the dependence of graph colorability on the nature and level 
of its connectivity, and the phase transitions that take place. The suggested 
framework comes with its own limitations; we critically discuss what can, and 
cannot be calculated via the methods of statistical mechanics.

The statistical physics approach is based on the introduction of a
Hamiltonian or cost-function, and the calculation of the typical free
energy in the large system limit. From the free energy one can obtain
the typical ground state energy, which in turn allows one to make
predictions on the graph colorability. A non-zero ground state
energy indicates that, under the given conditions, random graphs are
typically not colorable.  Our theoretical results are restricted to
the replica symmetric (RS) approximation (see~\cite{MPV,nishimori_book}), and
are, for the 2-color problem (which is solvable in linear time) in perfect
agreement with those obtained by numerical methods; for the 3-color problem the
results are only in qualitative agreement with those obtained by numerical
methods. The theoretical results can be systematically improved by using replica
symmetry breaking (RSB) approximations, although our current results do not
provide a direct indication for a breakdown of the RS approximation.

Apart from determining merely the colorability of the graph, the ground state 
energy also tells us what is the typical minimal fraction of unsatisfied edges 
when the graph is non-colorable. Furthermore, the ground state (residual)
entropy gives us information about the number of different coloring schemes
that share the minimum number of unsatisfied edges.

The suggested framework covers a range of possible  variations of the original
problem. However, only a limited number of them can be studied in a single
paper; we therefore restrict this study to regular $p=2$ and $p=3$ color
problems on random graphs with 2-vertex edges  (i.e. with 2-body interactions in
the statistical physics terminology). In  this context, {\em regular} stands for
the fact that all edges connect the same number of vertices and impose the same
color constraint on the vertices they connect, and that all vertices have the
same available color set.
Possible variations include many-$K$ vertex edges ($K$-body interactions);
mixtures of edges with different $K$ values and/or with different local
constraints imposed on the colors of the vertices involved; constraints on the
overall frequencies of vertices of a certain color; mixtures of vertices with
different available color sets; other probability distributions of the number of
edges per vertex, etc.

Our results, in agreement with results obtained elsewhere~\cite{Walsh}, seem
to indicate, that for $p\geq3$ there is a 1st order transition for the 
colorability as a function of the average graph connectivity, from probability 1
to 0, at some critical average connectivity. This implies that in 
these cases a vanishing ground state energy implies that the graph is 
$p$-colorable, while a non-zero ground state energy indicates that the graph is
typically not $p$-colorable.

Contrasting results obtained from the theoretical framework with numerical
studies in the case of $p=2$ exposes inherent limitations of the statistical 
physics based analysis. Using a completely different approach, we also obtain 
an exact expression for the probability that large random graphs with 2-vertex 
edges are 2-colorable, finding a 2nd order transition for the colorability as 
a function of the graph's average connectivity, from non-zero to zero
probability. This result shows that in general a zero ground state energy does
not automatically imply that a graph is typically colorable.

The paper is organized as follows: In section~\ref{sec:DN} we define
the problem and introduce the notations used, while in 
section~\ref{sec:RS} we introduce the statistical physics framework.
Section~\ref{sec:RE} introduces the results obtained from the analysis
as well as numerical results obtained by exact enumeration and Monte-Carlo
simulations. The case of 2-colorability is studied in 
section~\ref{sec:2E}; discussion and conclusions are presented in 
section~\ref{sec:CO}.

%
\section{Graph Coloring - Definitions and Notation \label{sec:DN}}
In a general set-up, we consider {\em regular} random graphs $\cG(N_v,K,\la)$
consisting of $N_v$ vertices, connected to each other by (hyper)
edges. Each (hyper) edge connects exactly $K$ distinct vertices. The
connectivity is then described by the tensor $\cD_{\{j_1..j_K\}}$, the elements
of which are $1$ is there is an (hyper) edge connecting the vertices
$\{j_1..j_K\}$, and 0 otherwise. Note that the total number of possible (hyper)
edges in the graph is given by $N_{pe/g}=\vtd{N_v}{K}$, while the total number
of possible (hyper) edges a given vertex $j$ may be involved in, is given by
$N_{pe/v}=\vtd{N_v\m1}{K\m1}$.
The overall connectivity of the graph $\cG(N_v,K,\la)$ is described by the
parameter $\la$, which gives the average number of edges each vertex is
involved in. Hence, for large graphs (i.e. $N_v\to\infty$), the fraction of the
total number of edges $N_e$ and the total number of vertices $N_v$, is
typically given by
\beq
{N_e\ov N_v}={\la\ov K}+\cO(N_v^{\m\ha}) \ .
\label{ev}
\eeq
In a {\em random} graph, this is obtained by considering all (i.e. 
$N_{pe/g}$) possible $K$-tuples $\{j_1..j_K\}$ of vertices, and by assigning
\beq
 \cD_{\{j_1..j_K\}}=
\lc\bay{ll} 1&{\rm with~probability}~P_e ~~=~~~~{\la\ov N_{pe/v}}\\
            0&{\rm with~probability}~P_{ne}=1-{\la\ov N_{pe/v}} \ .
\eay\rp
\label{D}
\eeq
In the large system limit (i.e. $N_v\to\infty$), the number of edges per vertex
($n_e$) is then Poisson distributed:
\beq
P(n_e=k)=
\vtd{N_{pe/v}}{k}\lh{\la\ov N_{pe/v}}\rh^k\lh1-{\la\ov N_{pe/v}}\rh^{N_{pe/v}\m 
k}\simeq {\la^k\ov k!}\exp(-\la),~~~~k=0,1,2,..,\infty \ .
\label{Pne}
\eeq
The most studied case is that of $K=2$, in which one considers conventional
edges (or 2-body interactions); graphs with $K\geq3$
are also considered in other contexts, for instance, the assignment of 
examination rooms to classes~\cite{graphcoloring}, in which case one generally
speaks of $K$ hyper edges (or $K$-body interactions).
Although we will derive expressions for general $K$, in this paper we will 
limit ourselves to the analysis of random graphs with $K=2$.

Now we assume that each vertex $j$ can take a color $c_j$ out of a set
$\{\mu_{j,1},..,\mu_{j,p_j}\}$ of $p_j$ colors, its {\em color set}.
A coloring problem on a graph is determined by the constraint(s) on the colors
of vertices connected by a(n) (hyper) edge. For instance, one can demand
that none of the colors $c_j$ of the vertices connected by an edge are the 
same, or that the colors $c_j$ of the vertices connected by an edge are not 
all the same (note that both constraints are identical for $K=2$). Although in 
principle, one can consider scenarios where the color set may differ
from vertex to vertex, and where the color constraints may differ from edge to
edge, in the present paper we restrict ourselves to the case where all vertices
have the same color set $\{\mu_1,..,\mu_p\}\ev\{1,..,p\}$ of $p$ colors, and
where each edge imposes the same color constraint on the vertices it connects.
The actual color of a vertex $j$ is indicated by $c_j\in\{1,..,p\}$, and we
denote a {\em coloring} of the entire graph by $\vcc\ev\{c_1,..,c_{N_v}\}$.

In this context our goal is to determine the probability that a randomly
generated graph with average connectivity $\la$, and a given color set and 
color constraints, is colorable.

Note that the $K=2$ case with $p$ available colors, is exactly the
anti-ferromagnetic $p$-spin Potts model~\cite{antiferroPotts}, while
the $p=2$ case is the anti-ferromagnetic Ising model
(see~\cite{MPV,nishimori_book}). The only randomness present in the
model, is the random graph connectivity.

%
\section{Replica Calculation \label{sec:RS}}
%
\subsection{General Scenario}
%
We now present the statistical physics formulation of the graph coloring
problem. To map this problem onto a statistical physics framework, we introduce
a Hamiltonian or cost function for given coloring $\vcc$ and connectivity $\cD$:
\beq
\cH(\vcc,\cD)\ev 
\sum_\BKK\cD_\BKK~\chi_\BKK(\vcc) \ ,
\label{H}
\eeq 
where we have introduced the following short-hand
notation for the $K$-tuples to keep our notation concise:
\beq
\BKK\ev\{j_1..j_K\}  \ .
\label{BKK}
\eeq
Furthermore, $\chi_\BKK(\vcc)$ is 0 if the edge color constraints are
satisfied and 1 otherwise, such that $\cH(\vcc)$ counts the number of
unsatisfied edges.
We focus on the case where colors of nodes sharing an edge should not
all be the same; $\chi_\BKK(\vcc)$ is then given by
\beq
\chi_\BKK(\vcc)=\sum_{\mu=1}^p[\mu_c]_\BKK,\hsc{\rm with}\hsc
\mu_{c_j}\ev \de_{\mu,c_j},\hsc 
[\mu_c]_\BKK\ev\prod_{k=1}^K\mu_{c_{j_k}} \ , 
\eeq
such that
\beq
\exp\lh-\be~\chi_\BKK(\vcc)\rh=\prod_{\mu=1}^p\lv 1-\De~[\mu_c]_\BKK\rv
=1-\De\sum_{\mu=1}^p[\mu_c]_\BKK \ ,
\label{Xi}
\eeq
where $\De\ev(1-e^{-\be})$.
In addition we could put constraints on the total fraction's $f_\mu$
of edges of color $\mu$:
\beq \sum_{j=1}^N \mu_{c_j}=Nf_\mu,\hsc ({\rm note~that}~~ \sum_\mu
f_\mu=1) \ .
\label{con}
\eeq
The central quantity from which all other relevant physical quantities
can be derived, is the free energy. This can be obtained from the
partition function (with the constraints on $\vf$):
\beq
\cZ(\vf,\cD)=\Tr_{\vcc}
\exp\lh-\be\sum_\BKK\cD_\BKK~\chi_\BKK(\vcc)\rh
\prod_\mu\de(\sum_{j=1}^N \mu_{c_j}-Nf_\mu)  \ .
\label{Z}
\eeq
The free energy per degree of freedom, is then obtained from
$\cF(\vf,\cD)=-{1\ov\be N_v}\log[\cZ(\vf,\cD)]$. It is very hard and not very
useful to calculate $\cF(\vf,\cD)$ for any specific choice of connectivity
$\cD$. Therefore, we calculate the expectation (average) value of the free
energy over the ensemble of all allowed realizations of the connectivity. The
average over all tensors $\cD$ with $K$ non-zero elements per row, and $L_j$ per
column $j$ is given by
\beq
\Bra g(\cD )\Ket _{\cD }\ev 
{\Tr_\cD g(\cD)\prod^{N_v}_{j_1=1}\de\lh\sum_{\Bra j_2,..,j_K\Ket}
     \cD_\BKK,L_{j_1}\rh \ov
 \Tr_\cD       \prod^{N_v}_{j_1=1}\de\lh\sum_{\Bra j_2,..,j_K\Ket}
     \cD_\BKK,L_{j_1}\rh}\ev {\cT\ov\cN}  \ .
\label{avD}
\eeq
Quantities of the type $\cQ(c)=\Bra\cQ_y(c)\Ket_y$, with
$\cQ_y(c)={1\ov M}\ln \left[{\cal Z}_y(c)\right]$ and
$\cZ_y(c)\equiv\Tr_{\!\!x}~f(x,y)$, are very common in the statistical
physics of disordered systems. We distinguish between the (quenched)
disorder $y$ (the connectivity $\cD$ in our case) and the microscopic
(thermal) variables $x$ (the coloring $\vcc$ in our case). Some
macroscopic order parameters $c(x,y)$ (the $f_\mu$ in our case) may be
fixed to specific values and may depend on both $y$ and $x$. Although
we will not prove this here, such a quantity is generally believed to
be {\em self-averaging} in the large system limit, i.e., obeying a
probability distribution $P(\cQ_y(c))= \de(\cQ_y(c)-\cQ(c)))$.  The
direct calculation of $\cQ(c)$ is known as a {\em quenched} average
over the disorder, but is typically hard to carry out, and requires
using the replica method~\cite{nishimori_book}. The replica method
makes use of the identity
$\Bra\ln\cZ\Ket=\Bra~\lim_{\rn\to0}[\cZ^\rn\m1]/\rn~\Ket$, by
calculating averages over a product of partition function
replicas. Employing assumptions about replica symmetries and
analytically continuing the variable $\rn$ to zero, one obtains
solutions which enable one to determine the state of the system.  We
now present only the definitions and final expressions for the
relevant physical quantities as obtained by the replica
calculation. For the technical details we refer to appendix~\ref{A}.
\nl
The order parameters that naturally occur in this calculation are
\beq
q^\BKm_\BKmu\ev\sum_{j=0}^NZ_j~[\mu_{c_j}]^\BKm_\BKmu,\hsc \rmm=0,1,..,\rn  \ ,
\label{qdef}
\eeq
and their definition is enforced by the introduction of the corresponding
Lagrange multipliers $\htq^\BKm_\BKmu$. Here we have introduced short hand
notations for the $\rmm$-tuples of replica indices and their corresponding
colors:
\beq
\BKm \ev\Bra a  _\ell~|~\ell=1,..,\rmm\Ket,\hsc
\BKmu\ev\{   \mu_\ell~|~\ell=1,..,\rmm  \},\hsc
[\mu_{c_j}]^\BKm_\BKmu\ev\prod_{\ell=1}^\rmm\de_{\mu_\ell,c^{a_\ell}_j}  \ .
\label{BKm}
\eeq
Note the difference in notation for the replica indices $<\!.\!>$,
which all have to be different, and for the colors $\{.\}$ for which
multiple occurrence of the same color is allowed.

Since all replicas are subject to the same disorder the corresponding
variables, depending on just one replica index, must be equivalent
(index independent): $\htf^a_\mu=\htf_\mu$, $q_\mu^a=q_\mu$ and
$\htq_\mu^a=\htq_\mu$. To proceed with the calculation, one needs to
assume a certain order parameter symmetry for $q^\BKm_\BKmu$ and their
conjugates $\htq^\BKm_\BKmu$, for $\rmm>1$.  The simplest ansatz is
that all replica $\rmm$-tuples ($\rmm=2,..,\rn$) with the same color
set $\BKmu$ are equivalent. This ansatz is called the replica
symmetric ansatz (RS). In RS the order parameters $q^\BKm_\BKmu$,
$\htq^\BKm_\BKmu$ depend only on the color multiplicities
$m_\mu\ev\sum_{\ell=1}^\rmm\de_{\mu,\mu_\ell}$ appearing in the $\rmm$-tuple
$\BKmu$ (i.e. $q^\BKm_\BKmu=q_\vm$, and $\htq^\BKm_\BKmu=\htq_\vm$,
where $\vm=\{m_\mu~|~\mu=1,..,p\}$).  Note that for general positive
integer $\rn$ there may be $\rmm$-tuples of any size up to $\rn$,
therefore $m_\mu$ can take the values $0,1,..,\rn$ under the
constraint $\sum_\mu m_\mu\leq\rn$~.  To facilitate the analytic
continuation to non-integer $\rn$, it is now technically advantageous
to write the discrete set of order parameters $\{q_\vm, ~\htq_\vm\}$
as the moments of $p$-variable probability distributions on the interval
$[0,1]^p$:
\beq
\lc\begin{array}{lll}   
   q_\vm=&   q_0\int'\{d \vx~  \pi( \vx)\}~\prod_{\mu=1}^p(x_\mu)^{m_\mu}   \\\\
\htq_\vm=&\htq_0\int'\{d\vhx~\htpi(\vhx)\}~\prod_{\mu=1}^p(-\htx_\mu)^{m_\mu}
\end{array}\rp \ ,
\label{RS}
\eeq
where $\int'd\vec{y}...\ev\int_0^1\{\prod_{\mu=1}^pdy_\mu\}...~
\de(\sum_{\mu=1}^py_\mu-1)$. The variables $x_\mu$ can be interpreted as the
(cavity) probabilities that a vertex takes the colors $\mu\in\{1,..,p\}$, and
$\pi(\vx)$ is their joint probability distribution. The constraint
$\sum_\mu y_\mu=1$ expresses the fact that the total probability is 1. Using the
ansatz (\ref{RS}), solving the saddle point equations with respect to $\htq_0$
and $q_0$, and taking the limit $\rn\to0$, we obtain the quenched free energy
per edge $\cF_e(\vf)$ for given values $\vf$:
\beq
\cF_e(\vf)={1\ov\be}\lv{K\ov\la}\sum_{\mu=1}^pf_\mu\htf_\mu+K\cG_1-\cG_2
-{K\ov\la}\cG_3\rv \ , 
\label{FRS} 
\eeq 
taken in the extremum with respect to $(\vhf,\htpi,\pi)$, where
\bea
\cG_1&\ev&\int'\{d\vx d\vhx~\pi(\vx)~\htpi(\vhx)\}
\log(1-\sum_{\mu=1}^p x_\mu\htx_\mu)                                       \nn\\
\cG_2&\ev&\int'\prod_{k=1}^K\{d\vx_k~\pi(\vx_k)\}
\log(1-\De\sum_{\mu=1}^p\prod_{k=1}^Kx_{k,\mu})                            \nn\\
\cG_3&\ev&\sum_{L=0}P(L)\int'\prod_{l=1}^L\{d\vhx_l~\htpi(\vhx_l)\}
\log\lh\sum_{\mu=1}^p\exp(\htf_\mu)\prod_{l=1}^L(1-\htx_{l,\mu})\rh \ .
\label{GRS}
\eea
The internal energy and entropy per edge are then given by
\beq
\cU_e={\pa\be\cF_e\ov\pa\be}=
\int'\prod_{k=1}^K\{d\vx_k~\pi(\vx_k)\}      
{\exp(-\be)\sum_{\mu=1}^p\prod_{k=1}^Kx_{k,\mu}\ov
 (1-   \De \sum_{\mu=1}^p\prod_{k=1}^Kx_{k,\mu})  },\hsc
\cS_e=\be(\cU_e-\cF_e)  \ .
\label{US}
\eeq
Note that it is convenient to consider the energy per edge ($\cU_e$),
and entropy per vertex ($\cS_v\ev{K\ov\la}\cS_e$). In this way,
$\cU_e$ is just the fraction of unsatisfied edges
(i.e. $0\leq\cU_e\leq1$), while $\cS_v$ is the entropy per degree of
freedom (i.e. $0\leq\cS_v\leq\log(p)$).

The saddle point equations are obtained by variation with respect to
$\vhf$, $\pi$ and $\htpi$ (under the constraint that $\pi$ and $\htpi$
are normalized) respectively, yielding:
\bea
  f_\mu    &=&\sum_{L=0}P(L)\int'\prod_{l=1}^L\{d\vhx_l~\htpi(\vhx_c)\}
  {              \exp(\htf_\mu)\prod_{l=1}^L(1-\htx_{l,\mu})\ov
   \sum_{\nu=1}^p\exp(\htf_\nu)\prod_{l=1}^L(1-\htx_{l,\nu})   }
\label{df}                                                                    \\
\htpi(\vhx)&=&\int'\prod_{k=1}^{K\m1}\{d \vx_k~\pi  ( \vx_k)\}
              \prod_\mu\de\lh\htx_\mu-\De\prod_{k=1}^{K\m1}x_{k,\mu}\rh
\label{dpi}                                                                   \\
  \pi( \vx)&=&\sum_{L=1}P(L){L\ov\la}
\int'\prod_{l=1}^{L\m1}\{d\vhx_l~\htpi(\vhx_l)\}\prod_\mu
\de\lh x_\mu-{      \exp(\htf_\mu)\prod_{l=1}^{L\m1}(1-\htx_{l,\mu})\ov
            \sum_\nu\exp(\htf_\nu)\prod_{l=1}^{L\m1}(1-\htx_{l,\nu})}\rh \ .
\label{dhpi}
\eea
From~(\ref{df}) and (\ref{dhpi}) we see that the normalizations $\sum_\mu
f_\mu=1$ and $\sum_\mu x_\mu=1$ are automatically satisfied.

Note that for the 2-color problem ($p=2$) one can invoke an Ising spin
representation for the colors, e.g. by mapping the color 1 onto spin
$\e1$ and color 2 onto spin $\m1$. Then, using the fact that
$x_2=1-x_1$, and defining $m\ev1-2x_2~(\in[\m1,1])$, one obtains a
single 1-variable probability distribution $\tilde\pi()$ for the
cavity {\em magnetization} ($m$) of the vertices (spins):
\beq
\tilde{\pi}(m)\ev\pi\Bigl( {1\e m\ov2},{1\m m\ov2} \Bigr)  \ .
\label{Pm}
\eeq
We also note that in the absence of overall color constraints
(i.e. $\htf_\mu=0$), a {\em paramagnetic} solution of the saddle point
equations (\ref{dpi},\ref{dhpi}) always exists:
\bea
  \pi_{\rm pm}(\vx )&=&\de\lh\vx -\Bigl( {1  \ov p       }\Bigr)\vec{1}\rh,\hsc 
\htpi_{\rm pm}(\vhx)=\de\lh\vhx-\Bigl({\De\ov p^{K\m1}}\Bigr)\vec{1}\rh   
\label{pmpi}                                                                \\
\cF_{e,\rm pm}&=&{1\ov\be}\lv{(K\la\m
K\m\la)\ov\la}\log(p)-\log(p^{K\m1}\m\De)\rv ,\hsc \cU_{e,\rm
pm}={\exp(-\be)\ov(p^{K\m1}\m\De)} ,\hsc f_\mu={1\ov p}
\label{pmFU}
\eea
Finally, one should note that the expressions (\ref{FRS}-\ref{dhpi}) are valid
for any distribution of the number of edges per vertex $P(L)$, although in this
paper we only investigate the case where $P(L)$ is a Poisson distribution.
\subsection{Two-body interactions, no color constraints}
%
We now derive explicit expressions for the special cases that we analyze in more
detail later on: 2-body edges, $K=2$, and no constraint on the overall color
frequencies $\htf_\mu=0,~\forall\mu$). From (\ref{dpi}) we obtain the relation
\beq
\htpi(\vhx)={1\ov\De}\pi\Bigl({\vhx\ov\De}\Bigr)\hsc\to\hsc
\int' d\vhx~\htpi(\vhx)~g(\vhx)=\int' d\vx~\pi(\vx)~g(\De\vx),
\eeq
such that the free energy per edge can be written in terms of the 
$p$-dimensional probability distribution $\pi(\vx)$ alone:
\bea
\cF_e&=&{1\ov\be}\lv\cG_1-{2\ov\la}\cG_3\rv,                                  \\
\cG_1&=&\int'\prod_{k=1}^2\{d\vx_k~\pi(\vx_k)\}~
        \log\lh1-\De\sum_\mu \prod_{k=1}^2 x_{k,\mu}\rh                       \\
\cG_3&=&\sum_L P(L)\int'\prod_{l=1}^L\{d\vx_l~\pi(\vx_l)\}\log\lh\sum_\mu 
        \prod_{l=1}^L (1-\De x_{l,\mu})\rh                  \ .                  
\eea
The saddle point equation (\ref{dhpi}) now becomes
\beq
\pi( \vx)=\sum_{L=1}P(L){L\ov\la}\int'\prod_{l=1}^{L\m1}\{d\vx_l~\pi(\vx_l)\}
          \prod_\mu\de\lh x_\mu-{     \prod_{l=1}^{L\m1}(1-\De x_{l,\mu})\ov
                         \sum_\nu     \prod_{l=1}^{L\m1}(1-\De x_{l,\nu})}\rh \ .
\label{dhpi2}
\eeq
Since the main question we want to investigate is the colorability of the graph,
we are specifically interested in the ground state energy. We therefore take the
low temperature limit (i.e. $\be\to\infty$), where a finite contribution to the
energy only exists when $1-x_{k,\mu}\ev\eps_{k,\mu}=\cO(\exp(-\be))$ for the
same color $\mu$ for both $k=1,2$; i.e. when two connected vertices are forced
to have the same color. Then the integrand of ({\ref{US}) becomes to leading
order,
\beq
{\exp(-\be)\sum_{\mu=1}^p\prod_{k=1}^2x_{k,\mu}\ov
 (1-   \De \sum_{\mu=1}^p\prod_{k=1}^2x_{k,\mu})  }=
{(1-X)\ov(1+\De X\exp(\be))}\simeq{1\ov (1+\exp(\be)X)}=\cO(1),               
\label{U0}
\eeq
with
\beq
X\ev \eps_{1,\mu}+\eps_{2,\mu}-\eps_{1,\mu}\eps_{2,\mu}-
           \sum_{\nu\neq\mu}x_{1,\nu}x_{2,\nu}\simeq\cO(\exp(-\be)) \ .
\label{X}
\eeq
However, the limit $\be\to\infty$ is not easily taken analytically for the fixed
point equation~(\ref{dhpi2}). As we show in appendix~\ref{B}, even in this
limit, the extremizing distribution $\pi(\vx)$ is non-trivial, and we have not
found a way to obtain it analytically. We therefore solve equation~(\ref{dhpi2})
numerically to obtain the equilibrium distribution $\pi(\vx)$ which is in turn
used to obtain $\cF,~\cU$ and $\cS$.

The various integrations in the saddle point equations and the
resulting physical quantities are obtained by the Monte-Carlo
method. The distribution $\pi(\vx)$ is obtained as the
($p$-dimensional) histogram of a large population of size $N_P$ of
$p$-dimensional points $\{\vx_i~|~i=1,..,N_p\}$. All results presented
in this paper have been obtained using $N_P=10^6$.  The fixed point
equation~(\ref{dhpi2}) can then be solved by randomly updating
(i.e. replacing) one of the $\vx_i\to\vx_i'$. The update of $\vx_i'$
is carried out by, first, randomly picking a value $L$ with
probability $P(L){L\ov\la}$, then randomly picking $L\m1$
$\vx_{i_l}$'s, and finally using the r.h.s. of the arguments of the
$\de$-function in (\ref{dhpi2}) to calculate the resulting components
of $\vx_i'$.  This process is repeated until the histogram reaches a
steady state. Once this histogram is obtained, it can be used to
calculate the various physical quantities in similar fashion.

Note that in order to reach a sufficient numerical precision in the
low temperature limit for the components of the $\vx_i$, we either
save $x_{i,\mu}$ if $x_{i,\mu}\leq0.5$ or $\eps_{i,\mu}\ev1-x_{i,\mu}$
if $x_{i,\mu}>0.5$. This avoids precision loss, e.g. in calculating
$(1\m\De x_{i,\mu})$, when $x_{i,\mu}$ is very close to 1. Similar
steps are taken to keep sufficient numerical precision for the r.h.s
of the saddle point equation~(\ref{dhpi2}).

Furthermore, it should be noted that often a very large number of iterations is
needed (up to $10^3N_P$) before the distribution becomes stationary. This, in
combination with the finite population size $N_P=10^6$, and the inherent
randomness in the Monte-Carlo integrations, puts a limit on the achievable
numerical precision of our results.
%
\section{Results \label{sec:RE}}
%
We now turn to the results of the numerical evaluation of the RS
expressions.

First, it should be noted that the residual entropy $\cS_0(\la)$ per vertex
(i.e. the logarithm of the number of colorings of the ground state) does not
vanish for any finite $\la$. For the 2-color problem
\beq
\cS_0(\la)\geq{N_{dc}(\la)\ov N_v}\log(2)\geq P(n_e=0,\la)\log(2)\ev\cS_l(\la)
>0 \ ,
\label{S02}
\eeq
where $N_{dc}(\la)$ is the number of disconnected clusters, and where
$P(n_e=0,\la)>0$ is the fraction of completely isolated vertices at
given connectivity $\la$. For each of these clusters, one can pick a
single representative vertex and give it 2 different colors; the color
of all the other vertices in the cluster is then uniquely determined
when the graph is 2-colorable. In the case of non 2-colorable cluster,
there is at least 1 (and possibly more) way of coloring the
remaining vertices such that the number of unsatisfied edges in the
cluster is minimal.

For the $p$-color problem
\beq
\cS_0(\la)\geq\sum_{k=0}^{p\m2}P(n_e=k,\la)\log(p-k)\ev\cS_l(\la)>0 \ ,
\label{S0p}
\eeq
where $P(n_e=k,\la)>0$ is the fraction of vertices connected by $k$
edges at given connectivity $\la$. A vertex connected to $k$ other
vertices, can at least pick between $p-k$ colors (and more if some of
the vertices it is connected to have the same color) whether the graph
is $p$-colorable or not.  In case the graph is not $p$-colorable,
there is at least 1 (and possibly more) choices of coloring the
vertices such that the number of unsatisfied edges in the graph is
minimal.

The ground state energy $E_0(\la)$ per edge can then be used as an
indicator for the colorability of the graphs. Since we use the saddle
point method, there may be $\cO(1/\sqrt{N_v})$ fluctuations of the
internal energy per edge around the saddle point value. If
$E_0(\la)\geq0$, this clearly precludes colorability, while for
$E_0(\la)=0$ the colorability may depend on the fluctuations.

Note that in the absence of overall constraints on the color
frequencies, the solutions always exhibit a complete color symmetry, as
expected. In other words, the distribution $\pi(\vx)$ is symmetric
under permutations of the components of $\vx$ (up to numerical
precision), and the marginal distribution for each of the colors is
identical:
\beq
\tilde{\pi}_\mu(x_\mu)\ev\int_0^1 \prod_{\nu\neq\mu}^p dx_\nu~\pi(\vx),\hsc\to
\hsc\tilde{\pi}_\mu(x_\mu)=\tilde{\pi}(x)~~~,\forall~\mu
\eeq
%

\subsubsection{2-color graphs}
For the 2-color problem, the results are as follows (see Figs.\ref{fig:rp2} and
\ref{fig:pd2}):
\begin{itemize}
\item For $\la\leq1$, we only find the paramagnetic solution at all
temperatures and the corresponding ground state energy $E_0(\la)=0$.
\item For $\la>1$, from a certain (inverse) temperature $T_p(\la)$
($\be_p(\la)$) on-wards, the paramagnetic solution coexists with a non-trivial
solution, which can be identified as the physical one (at least in the RS
approximation) by the fact that this solution continues to obey inequality
(\ref{S02}) for all values of $\la$ that we have examined, while the
continuation of the para-magnetic solution violates it. We have a positive
ground state energy $E_0(\la)>0$, and in perfect agreement with the numerical
experiments, this predicts $P_c(\la)=0$ for $\la>\la_c=1$.
\end{itemize}

The behavior of the ground state energy and entropy is presented in
Fig.\ref{fig:rp2} while the phase diagram and the explicit
distribution obtained above $\la>1$ is presented in Fig.\ref{fig:pd2}.

>From (\ref{US}), we see that the internal energy is always positive.
Furthermore, the numerical analysis indicates that also the entropy and the
specific heat $C_V\ev{\pa\cU\ov\pa T}=T{\pa \cS\ov\pa T}$ are always
non-negative, and inequality~(\ref{S02}) is always satisfied. This implies that
all quantities behave as in a proper physical system, not giving any direct
indication that the RS-ansatz might be inaccurate.
\begin{figure}[h]
\setlength{\unitlength}{1mm}
\begin{picture}(160,55)
\put( 0  , 0  ){\epsfysize=55\unitlength\epsfbox{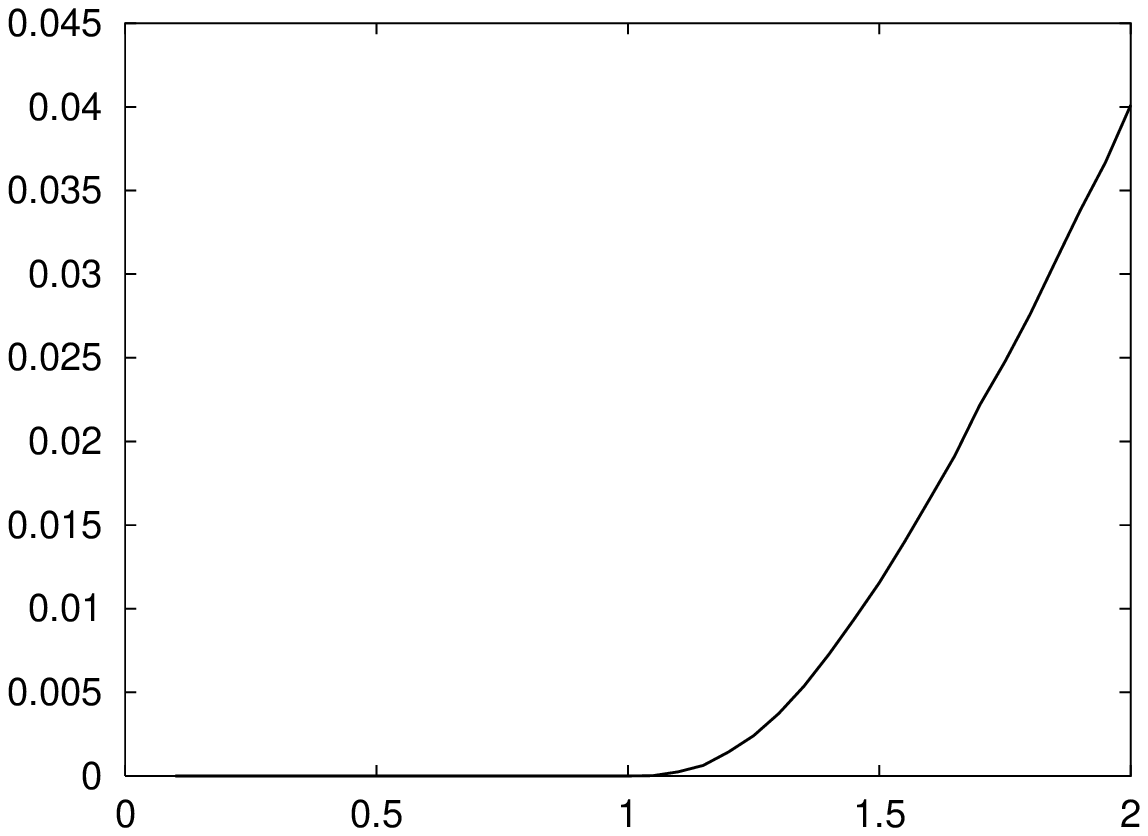}}
\put( 55 ,40  ){\mbox{\fns   \boldmath $E_0$}}
\put( 45 , 0  ){\mbox{\fns   \boldmath $\la$}}
\put(42.3, 4.3){\mbox{\tiny  \boldmath $\bullet$}}
\put( 15 ,48  ){\mbox{\fns   \boldmath $p=2$}}
\put( 80 , 0  ){\epsfysize=55\unitlength\epsfbox{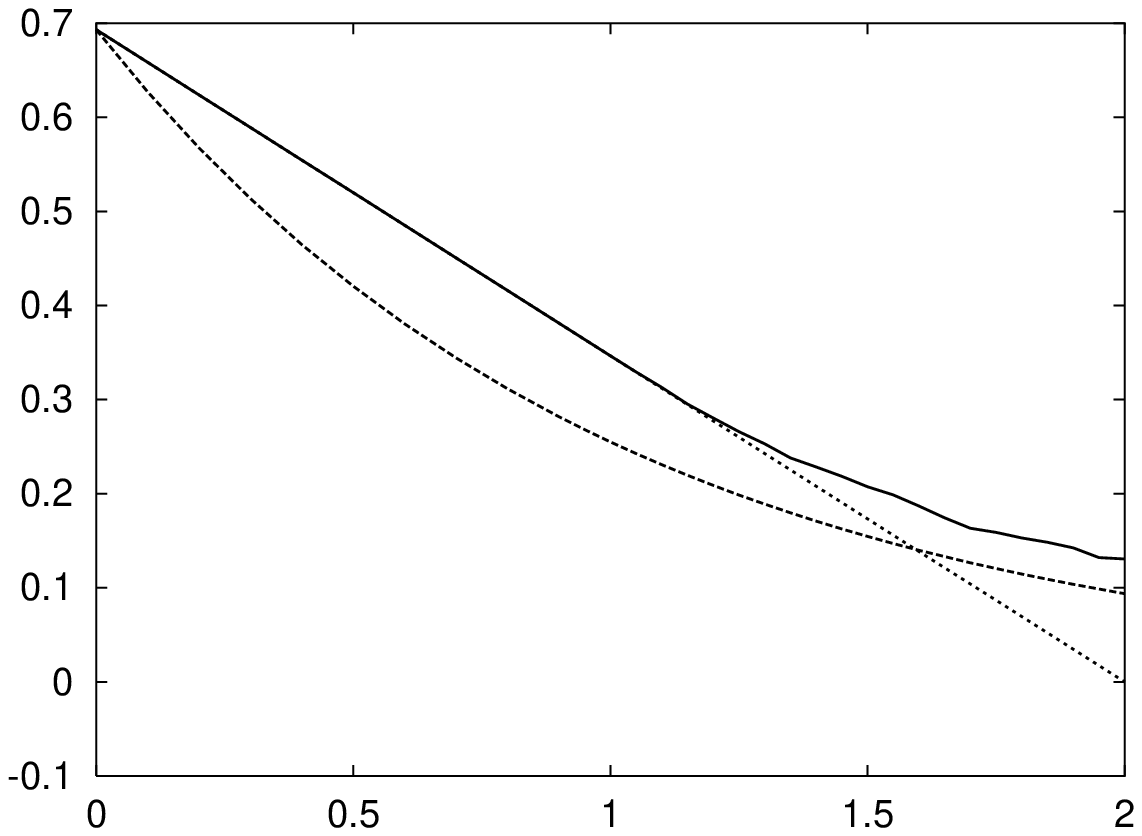}}
\put(140 ,25  ){\mbox{\fns   \boldmath $\cS_0$}}
\put(140 ,15  ){\mbox{\fns   \boldmath $\cS_{\rm pm}$}}
\put(110 ,27  ){\mbox{\fns   \boldmath $\cS_l$}}
\put(121.8,30.4){\mbox{\tiny \boldmath $\bullet$}}
\put(125 , 0  ){\mbox{\fns   \boldmath $\la$}}
\put(100 ,48  ){\mbox{\fns   \boldmath $p=2$}}
\end{picture}
\caption{On the left: the ground state energy $E_0(\la)$ for $p=2$. Up to
$\la=1$ ($\bullet$), $E_0(\la)=0$. The paramagnetic ground state energy
$E_{0,\rm pm}$ is always 0.  On the right: the ground state entropy $S_0(\la)$
(full line) for $p=2$, compared to its lower bound $S_l(\la)$~(\ref{S02})
(dashed line), and the paramagnetic ground state entropy $S_{\rm pm}(\la)$
(dotted line). Up to $\la=1$ ($\bullet$), $S_0$ and $S_{\rm pm}$ coincide.
\label{fig:rp2}}
\end{figure}
\begin{figure}[h]
\setlength{\unitlength}{1mm}
\begin{picture}(160,55)
\put( 0  , 0  ){\epsfysize=55\unitlength\epsfbox{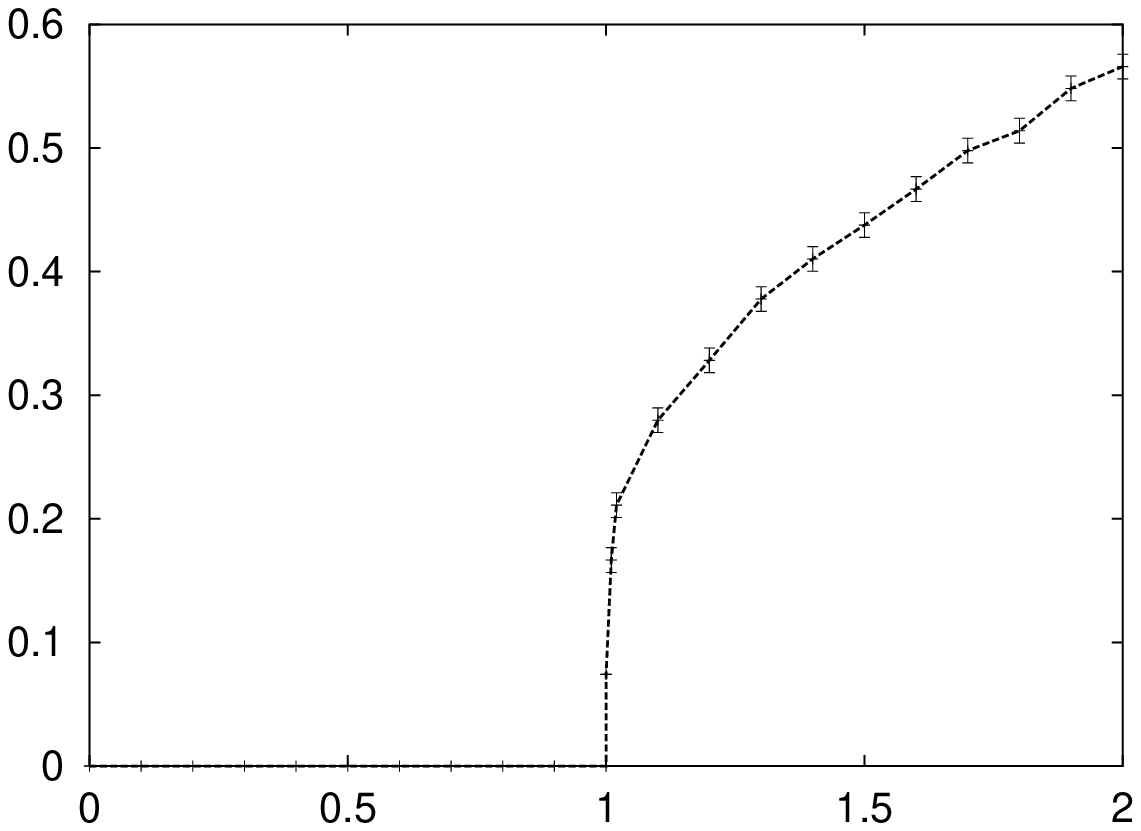}}
\put( 20 ,30  ){\mbox{\fns   \boldmath $PM$}}
\put( 55 ,20  ){\mbox{\fns   \boldmath $NPM$}}
\put(40.6,4.8  ){\mbox{\tiny  \boldmath $\times$}}
\put( 45 , 0  ){\mbox{\fns   \boldmath $\la$}}
\put(  0 ,40  ){\mbox{\fns   \boldmath $T  $}}
\put( 10 ,48  ){\mbox{\fns   \boldmath $p=2$}}
\put( 80 , 0 ){\epsfysize=55\unitlength\epsfbox{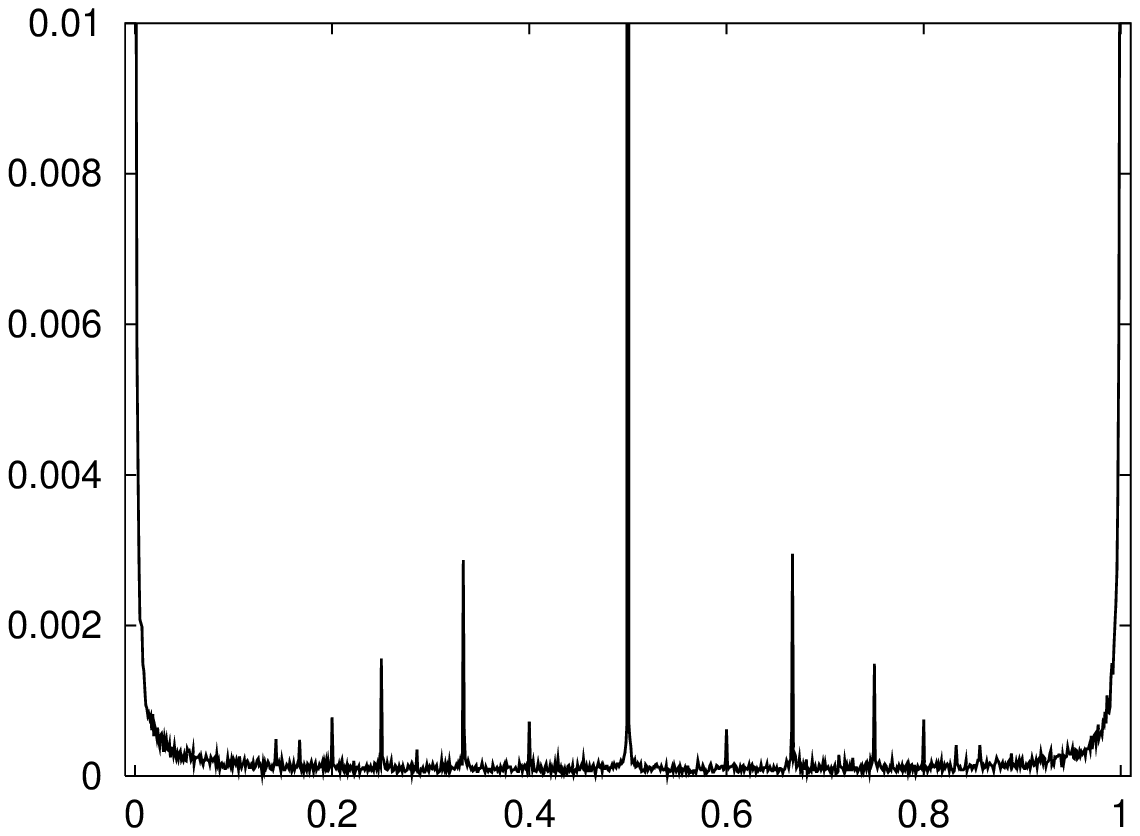}}
\put( 95 ,48  ){\mbox{\fns   \boldmath $p=2$}}
\put( 95 ,40 ){\mbox{\fns   \boldmath $\tilde\pi(x)$}}
\put(122 , 0 ){\mbox{\fns   \boldmath $x$}}
\end{picture}
\caption{On the left: the phase diagram $(\la,T)$, and the transition
from the paramagnetic to the non-paramagnetic RS state. The phase
transition is 2nd order in $\pi(\vx)$. At zero temperature, $E_0=0$
for $\la\leq1$ ($\times$) and $E_0>0$ for $\la>1$ (from $\times$ on-wards).
On the right: the stationary distribution $\tilde\pi(x)$ for $p=2$, $\la=2$
$(>\la_c)$, $\be=15$. We note the peaks and non trivial distribution at
$x\simeq0$ and $x\simeq1$, indicating that many vertices are forced (not) to
take a specific color. For $\la<\la_c$ these peaks are absent. We also note the
distinct peaks at $x\sim1/2,1/3$ and other rational values. The symmetry around
$x=0.5$ is specific for $p=2$.
\label{fig:pd2}}
\end{figure}

\subsubsection{3-color graphs}

For the 3-color problem, the results are as follows (see
Figs.\ref{fig:rp3} and \ref{fig:pd3}):
\begin{itemize}
\item For $\la\lessapprox 4$, we only find the paramagnetic solution at
all temperatures, and the corresponding ground state energy
$E_0(\la)=0$.
\item For $4\lessapprox\la\lessapprox 5.1$, from a certain temperature
$T_{pm}(\la)$ on-wards, the paramagnetic solution coexists with a
non-trivial solution, which can be identified as the physical one by
comparing the free energies. The ground state energy $E_0(\la)$
remains 0.
\item For $5.1\lessapprox \la$, from a certain temperature $T_{pm}(\la)$
on-wards, the paramagnetic solution coexists with a non-trivial solution with a
positive ground state energy $E_0(\la)>0$, which can be identified as the
physical one by by the fact that this solution continues to obey inequality
(\ref{S0p}) for all values of $\la$ that we have examined, while the
continuation of the para-magnetic solution violates it.
\end{itemize}

The behavior of the ground state energy and entropy is presented in
Fig.\ref{fig:rp3}; explicit distributions obtained for various $\la$
values are presented in Fig.\ref{fig:his3}, while the phase diagram
is presented in Fig.\ref{fig:pd3}.

As we will see, the numerical experiments predict that $P_c(\la)=1$
for $\la<\la_c\simeq 4.7$, and that $P_c(\la)=0$ for $\la>\la_c\simeq
4.7$.  Although the RS analysis results do not contradict the
numerical ones, they are unable to identify $\la_c\simeq 4.7$ as the
critical colorability value.  This is reminiscent of the RS results in
the K-SAT problem~\cite{SAT}.

In our case, however, from (\ref{US}), we see that the internal energy is always
non-negative. In addition , the numerical analysis shows that both entropy and
specific heat $C_V\ev{\pa\cU\ov\pa T}=T{\pa \cS\ov\pa T}$ are always
non-negative, and inequality (\ref{S0p}) is always satisfied.  This implies that
all quantities behave as in a proper physical system, thus giving no direct
indication that the RS-ansatz is wrong.

\begin{figure}[h]
\setlength{\unitlength}{1mm}
\begin{picture}(160,55)
\put( 0  , 0  ){\epsfysize=55\unitlength\epsfbox{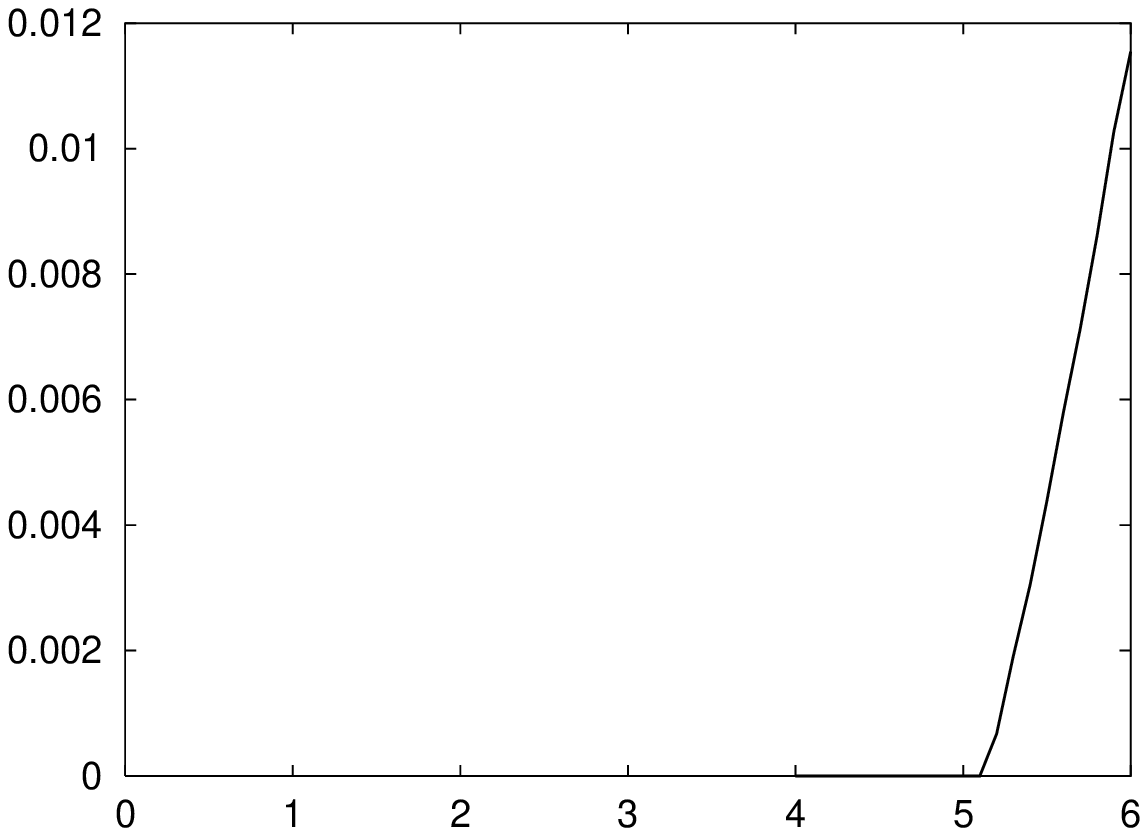}}
\put( 55 ,40  ){\mbox{\fns   \boldmath $E_0$}}
\put( 45 , 0  ){\mbox{\fns   \boldmath $\la$}}
\put( 64.1,4.2){\mbox{\tiny  \boldmath $\bullet$}}
\put( 15 ,48  ){\mbox{\fns   \boldmath $p=3$}}
\put( 80 , 0  ){\epsfysize=55\unitlength\epsfbox{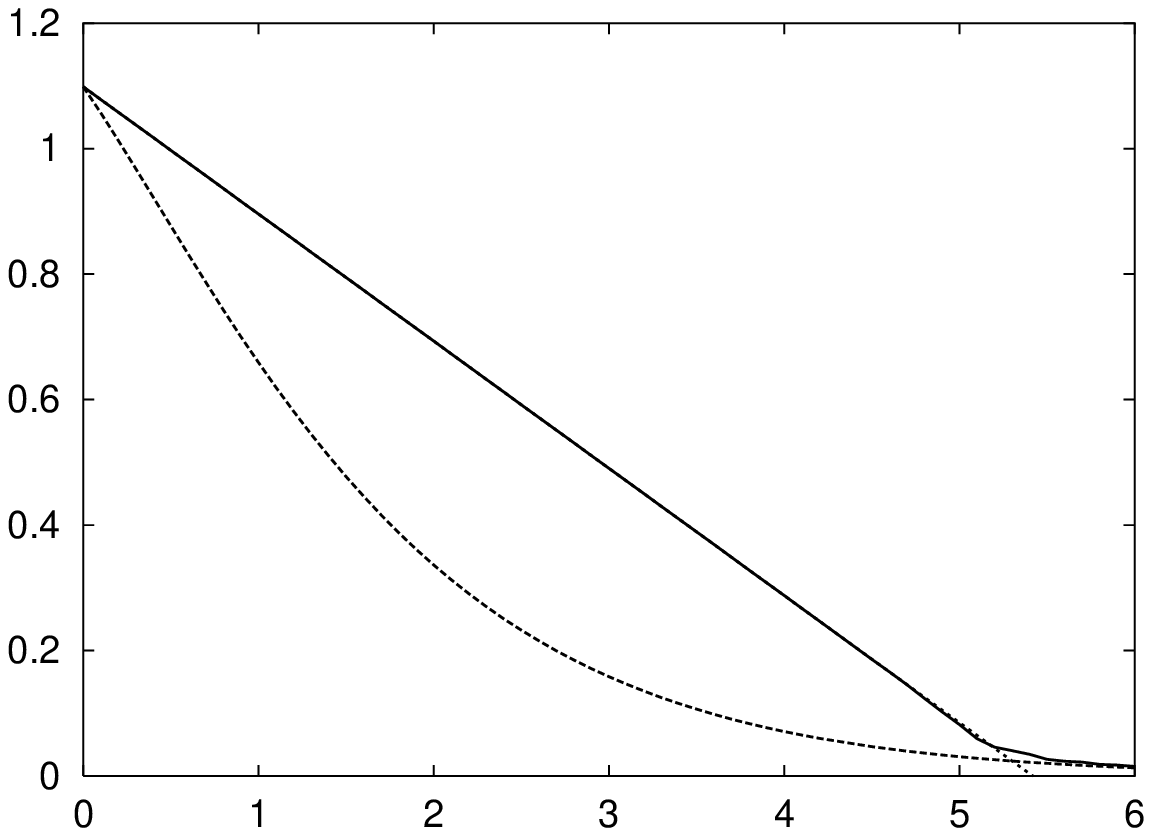}}
\put(132 ,10  ){\mbox{\fns   \boldmath $\cS_{\rm pm}$}}
\put(150 , 8  ){\mbox{\fns   \boldmath $\cS_0$}}
\put(105 ,27  ){\mbox{\fns   \boldmath $\cS_l$}}
\put(132 ,15.7){\mbox{\tiny  \boldmath $\bullet$}}
\put(125 , 0  ){\mbox{\fns   \boldmath $\la$}}
\put(100 ,48  ){\mbox{\fns   \boldmath $p=3$}}
\end{picture}
\caption{ On the left: the ground state energy $E_0(\la)$ for
$p=3$. Up to $\la\simeq5.1$ ($\bullet$), $E_0=0$. The paramagnetic
ground state energy $E_{0,\rm pm}$ is always 0.  On the right: the
ground state entropy $S_0(\la)$ (full line) for $p=3$, compared to its
lower bound $S_l(\la)$ (dashed line), and the paramagnetic ground state
entropy $S_{0,\rm pm}(\la)$ (dotted line). Up to $\la\simeq4$
($\bullet$), $S_0$ and $S_p$ coincide.
\label{fig:rp3}}
\end{figure}
\begin{figure}[h]
\setlength{\unitlength}{1mm}
\begin{picture}(160,55)
\put(  0 , 0 ){\epsfysize=55\unitlength\epsfbox{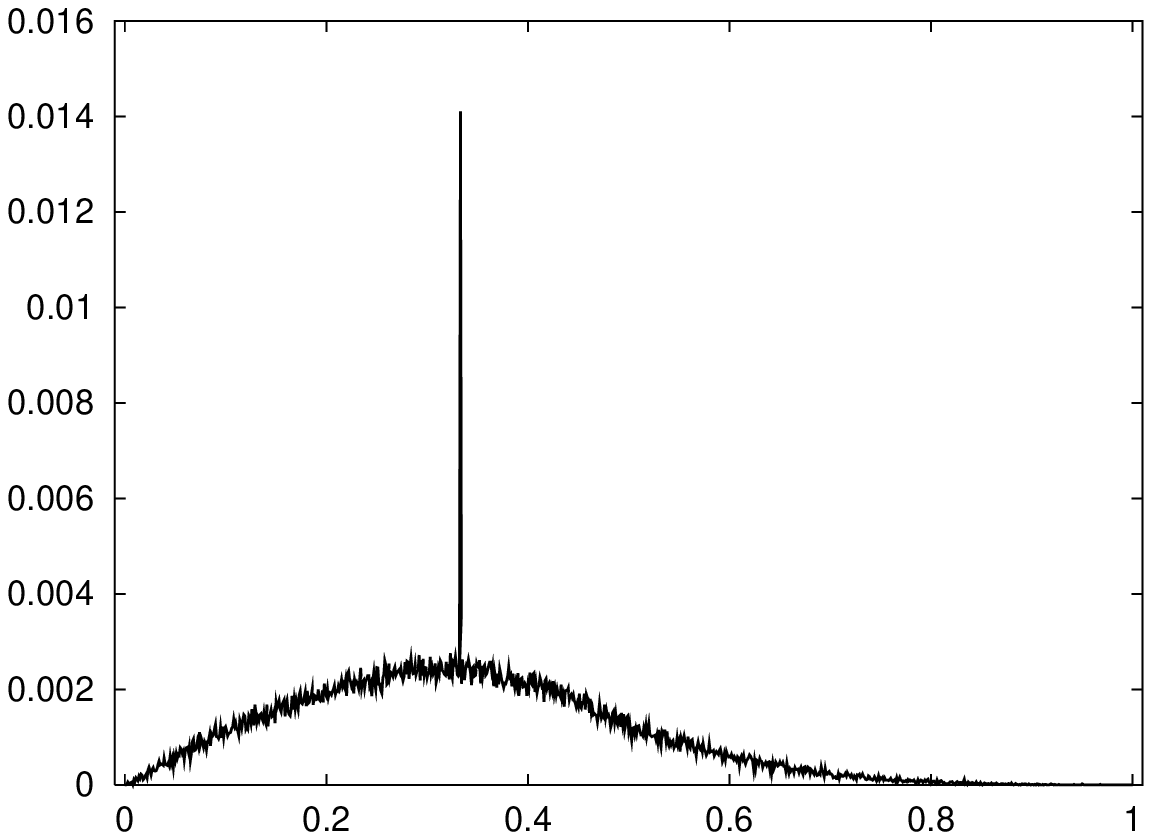}}
\put( 13 ,45 ){\mbox{\fns   \boldmath $\tilde\pi(x)$}}
\put( 41 , 0 ){\mbox{\fns   \boldmath $x$}}
\put( 35 ,20 ){\epsfysize=28\unitlength\epsfbox{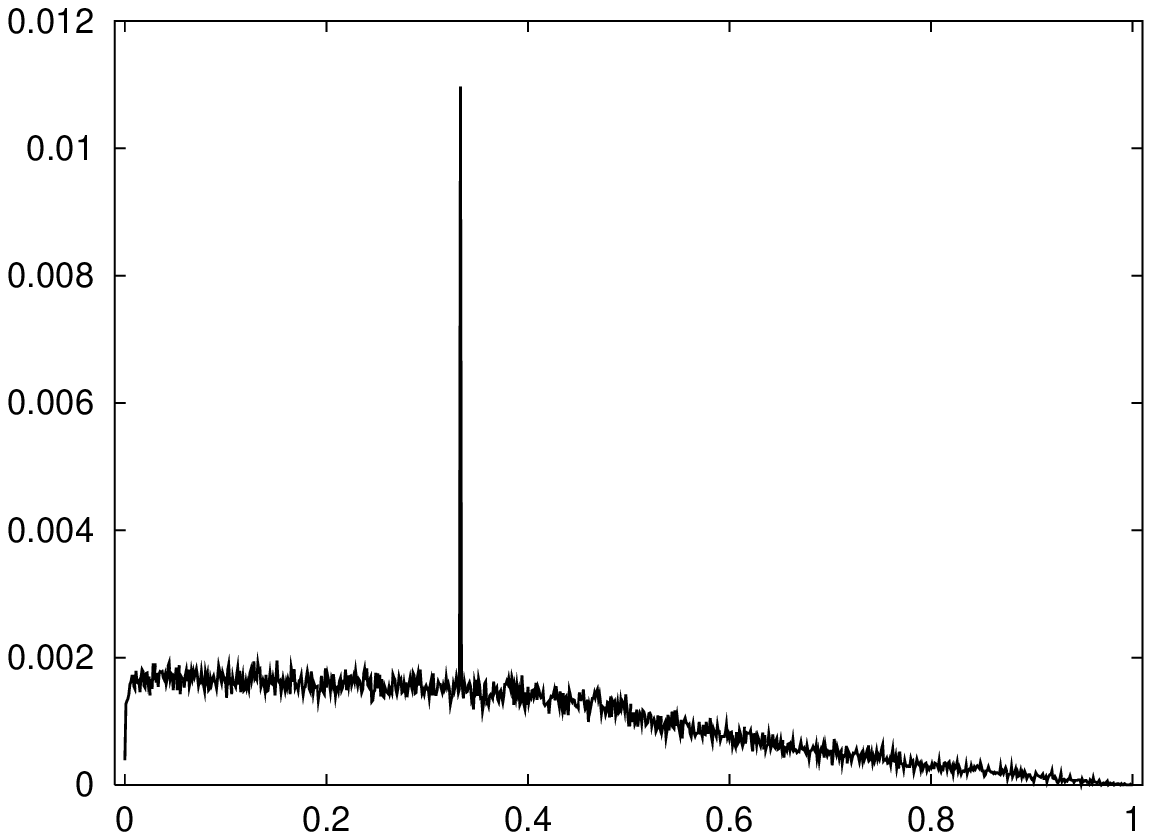}}
\put( 40 ,40 ){\mbox{\tiny   \boldmath $\tilde\pi(x)$}}
\put( 55 ,20 ){\mbox{\tiny   \boldmath $x$}}
\put( 80 , 0 ){\epsfysize=55\unitlength\epsfbox{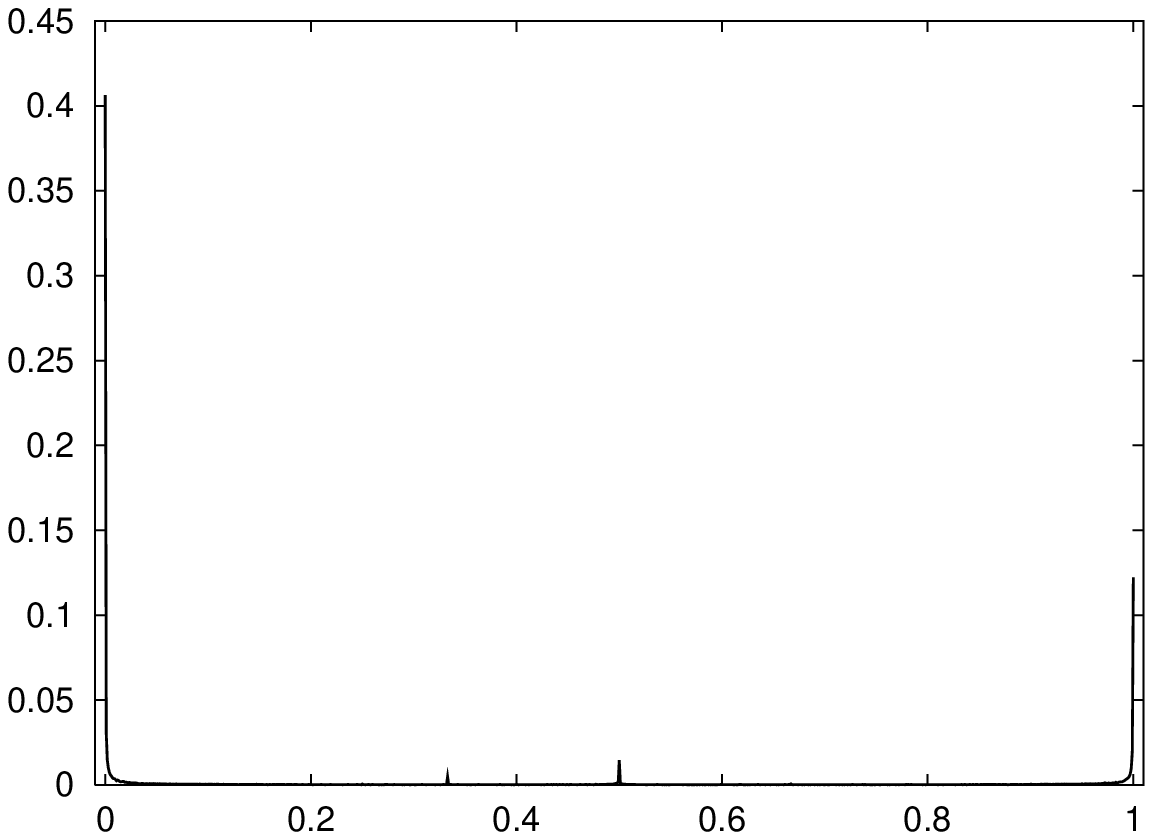}}
\put( 93 ,45 ){\mbox{\fns   \boldmath $\tilde\pi(x)$}}
\put(121 , 0 ){\mbox{\fns   \boldmath $x$}}
\put(100 ,15 ){\epsfysize=35\unitlength\epsfbox{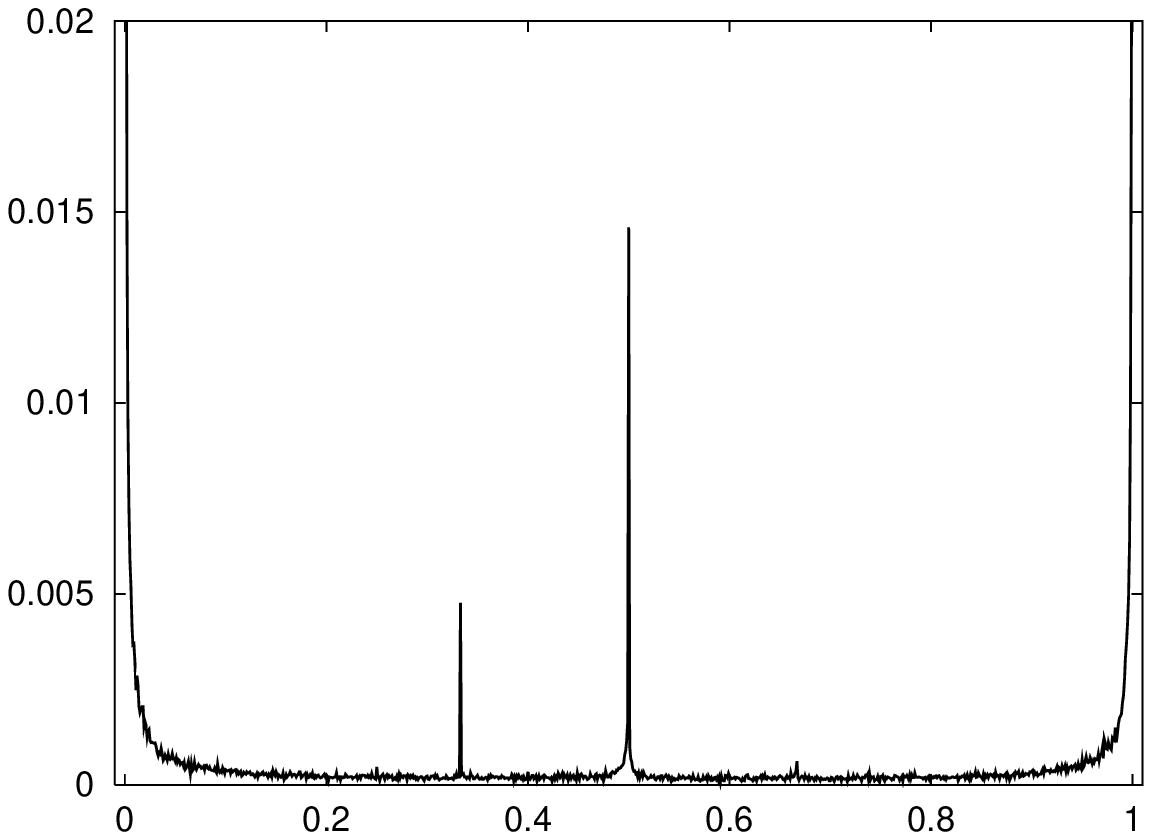}}
\put(112 ,45 ){\mbox{\tiny   \boldmath $\tilde\pi(x)$}}
\put(130 ,15 ){\mbox{\tiny   \boldmath $x$}}
\end{picture}
\caption{ On the left: the stationary distribution $\tilde\pi(x)$ for
$p=3$, $\la=4.5$ $(<\la_c)$ (and $\la=4.8$ inset), $\be=15$. Although
the solutions clearly differs from the paramagnetic solution (a single
peak at $x=1/3$), the absence of peaks near $x\simeq 0,1$ indicates
that $E_0(\la)=0$.  On the right: the stationary distribution
$\tilde\pi(x)$ for $p=3$, $\la=5.5$ $(>\la_c)$, $\be=15$. In the inset
we have enlarged and truncated the vertical scale, to illustrate the
continuous nature of the distribution.  We note the peaks and
non-trivial distribution at $x\simeq0$ and $x\simeq1$, indicating that many
vertices are forced (not) to take a specific color, and that $E_0(\la)>0$.
\label{fig:his3}}
\end{figure}
\begin{figure}[h]
\setlength{\unitlength}{1mm}
\begin{picture}(160,55)
\put( 50  , 0  ){\epsfysize=55\unitlength\epsfbox{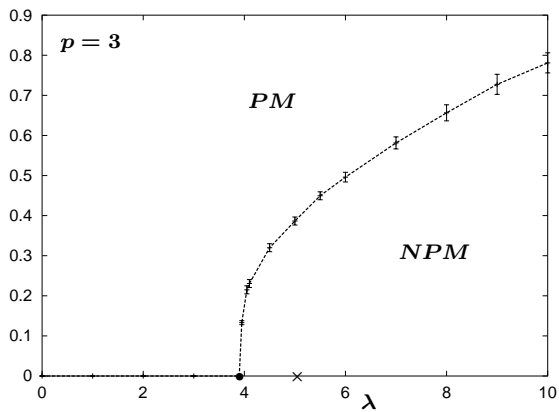}}
\put( 85  ,40  ){\mbox{\fns   \boldmath $PM$}}
\put(105  ,20  ){\mbox{\fns   \boldmath $NPM$}}
\put(100  , 0  ){\mbox{\fns   \boldmath $\la$}}
\put( 83.2, 3.8){\mbox{\tiny  \boldmath $\bullet$}}
\put( 90.5, 3.8){\mbox{\tiny  \boldmath $\times$}}
\put( 60  ,48  ){\mbox{\fns   \boldmath $p=3$}}
\end{picture}
\caption{The phase diagram $(\la,T)$. The phase transition from a
paramagnetic distribution $\pi(\vx)$ to a non-paramagnetic
distribution $\pi(\vx)$ is 2nd order in $\pi(\vx)$. From $\times$ on-wards the
RS ground state energy is positive.
\label{fig:pd3}} 
\end{figure}

%
\subsubsection{exact enumerations}
To validate the results obtained analytically we carried out extensive
computer simulations using two different approaches.

The first numerical method we use, is an exact enumeration of all the possible
colorings for a given graph. Note that, in general, the number of possible
colorings examined grows exponentially with the system size $N_v$ (i.e. $\sim
P(N_v)~\exp(cN_v\log(p\m1))$, where $P(N_v)$ is some polynomial, and where $c$
is some constant called the {\em attrition rate}, see e.g. \cite{EE} and
references therein. Hence, for $p\geq3$, we are fairly limited in the accessible
system sizes (i.e. $N_v\simeq\cO(10^2)$), and may expect considerable finite
size effects.

For $p=2$, however, the colorability of a graph can be determined by
the following linear algorithm: We start by picking a vertex at random, and 
giving it a certain color. Then, the color of all vertices it is connected to
(i.e. the 2nd generation, which is typically a finite number that depends on
$\la$), must have the opposite color, and the edges involved can be removed. Now
one can assign the first color to all the vertices (the 3rd generation)
connected to the 2nd generation, and the edges involved are again removed. This
process is repeated until either the whole graph is colored or a contradiction
is encountered. Since this process requires only a finite number of operations per
edge, and since the number of edges is $N_e={\la\ov 2}N_v$, one can determine
the 2-colorability of the graph in linear time, and large system sizes are
accessible. It is important to note that a graph that contains any loop of odd
length, is not 2-colorable, while any graph that does not contain a loop of odd
length, is. We will use this observation to obtain an exact expression for the
2-colorability of random graphs in the next section.

The 2-colorability $\cP_c(\la)$ as obtained by exact enumerations for system
sizes $N_v=10^2,..,10^5$, as well as the theoretical line (for $N_v\to\infty$),
are plotted in Fig.\ref{fig:eep}. We observe that $\cP_c(\la)$ decreases
continuously from $\cP_c(\la)=1$ at $\la=0$ to $\cP_c(\la)=0$ for
$\la\geq1$. These results are in full agreement with those reported
in~\cite{Walsh}, although here we have studied much larger systems. They are
also in agreement with the results obtained by the replica method, but the
latter is unable to distinguish between $\cP_c(\la)=1$ and $0<\cP_c(\la)<1$
as in both cases the ground state energy is 0. 

One should note that this linear algorithm is specific to the graph-coloring
problem with $p=2$ and $K=2$. In the case that $p\geq3$ and/or $K\geq3$, the
colors of the next generation are not uniquely determined by the colors of the
previous one. The same holds for the K-SAT problem (even with $K=2$) where a
clause (i.e. edge) may be satisfied by either of its arguments or by both.

For $p\geq3$ it is believed that no polynomial algorithm exists to
determine the $p$-colorability of a graph, and we have to resort to
the exploration of the possible colorings by building up a search
tree. Since we limit ourselves to determining whether a graph is
colorable or not, we are able to introduce some criteria to reduce the
problem, thus avoiding enumerating the full search tree.

A first step in the reduction is {\em pruning}: a vertex that has more
available colors than vertices it is connected to, will always be able
to satisfy all edges, irrespective of their colors.  Therefore, it
will not determine the colorability of the graph, and the vertex and
all its edges can be pruned. This pruning is to be done iteratively
(as the pruning of one vertex with its edges may render other vertices
prunable), until all remaining vertices have at least as many edges as
available colors.

A second step is {\em early stopping}: one starts coloring the
remaining vertices, keeping track of the remaining available colors
per vertex for all uncolored vertices. One can stop exploring the
search tree when a good coloring is found. Alternatively, when the
number of remaining available colors for a vertex becomes 0, the
coloring so far will lead to a contradiction later on, and we can
abandon this branch of the search tree altogether. One then backtracks
to the point where a coloring was still possible.

All this greatly reduces the actual number of colorings that have to
be examined, leaving it, however, exponential in the system size, thus
greatly limiting the accessible system size. Furthermore, since we
stop as soon as we encounter a contradiction, we have no information
on the minimum number of unsatisfied edges (i.e. the ground state
energy), or the number of colorings that yield the minimum number of
unsatisfied edges (i.e. the residual entropy).  In Fig.\ref{fig:eep}
we observe that the transition from $\cP(\la)=1$ to $\cP(\la)=0$
becomes increasingly sharp with the increasing system size, and that
the curves cross at $\la\simeq4.7$. This is typical for a 1st order
transition, and puts the critical connectivity for the infinite system
at $\la_c\simeq4.7$.  In this limit we expect $\cP(\la)$ to go
discontinuously from 1 to 0, in accordance with the results presented
in~\cite{Walsh}.
\begin{figure}[h]
\setlength{\unitlength}{1mm}
\begin{picture}(160,50)
\put(  0 , 0 ){\epsfysize=55\unitlength\epsfbox{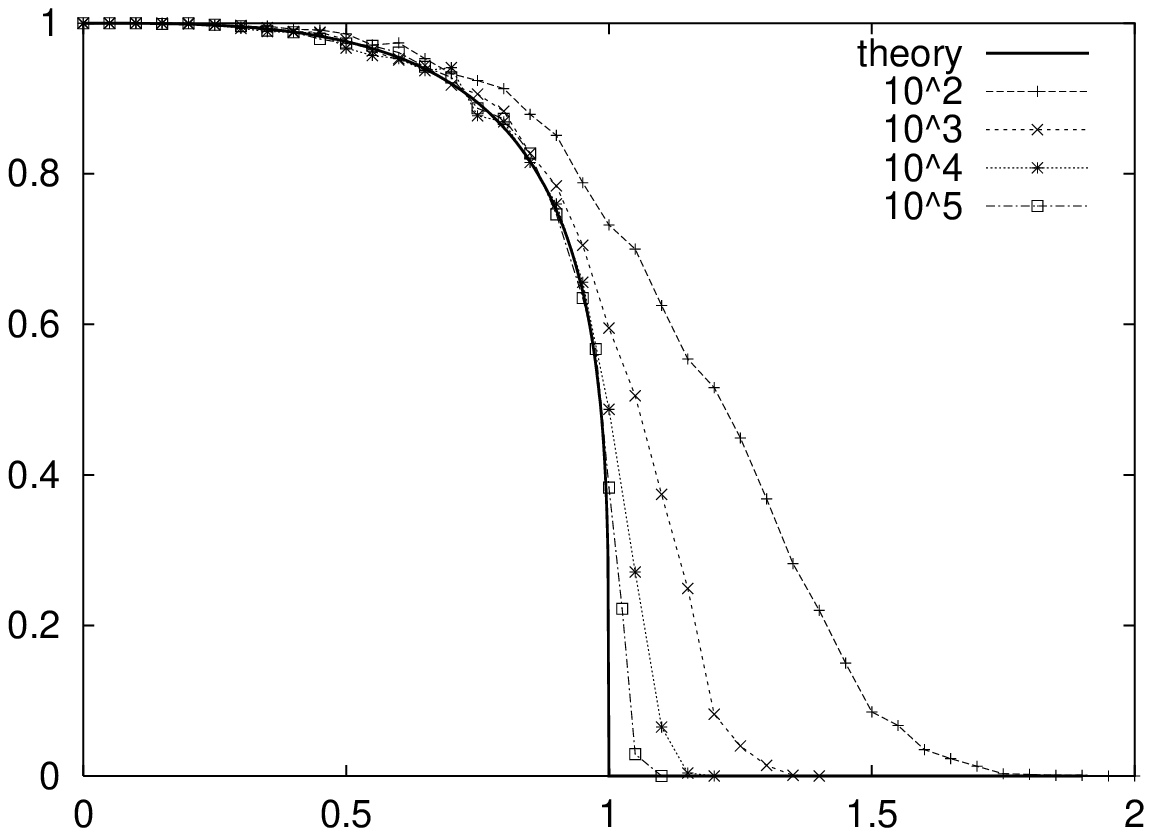}}
\put( 15 ,40 ){\mbox{\fns   \boldmath $\cP_c(\la)$}}
\put( 45 , 0 ){\mbox{\fns   \boldmath $\la$}}
\put( 80 , 0 ){\epsfysize=55\unitlength\epsfbox{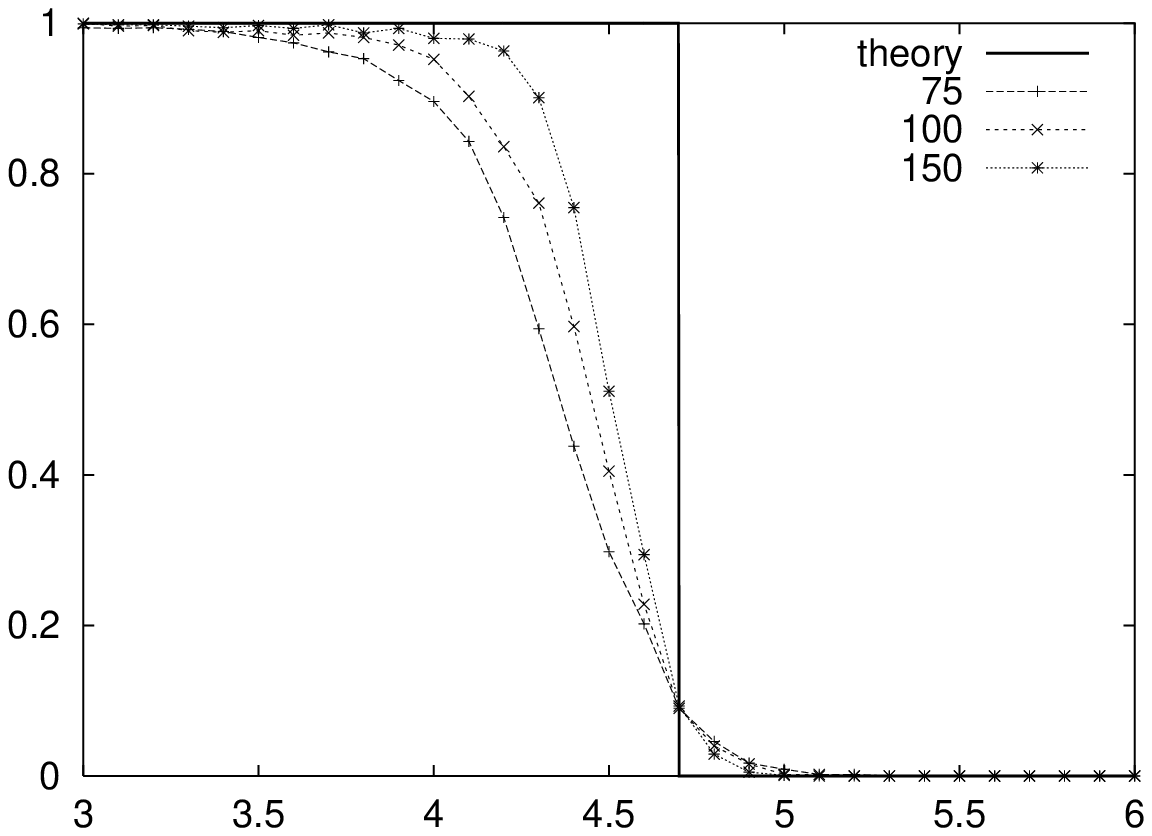}}
\put( 92 ,40 ){\mbox{\fns   \boldmath $\cP_c(\la)$}}
\put(125 , 0 ){\mbox{\fns   \boldmath $\la$}}
\end{picture}
\caption{ Left: the probability that a random graph is
2-colorable. The transition from $P_c(\la)>0$ to $P_c(\la)=0$ is 2nd
order.  Right: the probability that a random graph is 3-colorable. The
transition from $P_c(\la)>0$ to $P_c(\la)=0$ is 1st order.  The
probabilities are obtained by exact enumerations, averaged over $10^3$
runs.
\label{fig:eep}}
\end{figure}
%
\subsubsection{Monte-Carlo simulations}
%
Since exact enumerations for $p\geq3$ are limited to relatively small
system sizes, we have also performed Monte-Carlo simulations with
simulated annealing for the $p=3$ case. The simulations have been
performed for system sizes $N_v=1000$ and $N_v=10000$ and consist of
the following ingredients:
\begin{itemize}
\item At each temperature we perform Monte-Carlo dynamics. Starting
with a configuration $\vcc$ with energy $E(\vcc)$, we change the color
of a randomly chosen $c_j$ to $c'_j\neq c_j$, obtaining the new
configuration $\vcc'$ with energy $E(\vcc')$.  Then, if $\De E\ev
E(\vcc')-E(\vcc)\leq 0$ we always accept the move; otherwise we accept
it with probability $\exp(-\be~\De E)<1$.

\item We then gradually lower the temperature (this is known as simulated
annealing~\cite{SA}). If the temperature is reduced (cooling of the system)
logarithmically slowly with the system size, one is guaranteed to find the global
minimum $\vcc_0$ of $E(\vcc)$. However, logarithmically slow cooling is not
feasible due to limitations in computing time. Therefore, we must adopt a
feasible cooling scheme. Here we have opted for a linear cooling scheme, where
we increase $\be$ by small steps of fixed length $d\be=10^{-4}$. At each inverse
temperature $\be$ we make $C~N_v$ Monte-Carlo steps, and we control the cooling
rate by changing $C$, and try to extrapolate to $1/C\to0$ in order to obtain a
prediction for infinitely slow cooling. The values of $C$ that we have
considered, are $C=0.1,1,10,100$. The values of the ground state energy as
obtained by the linear cooling schemes serve as an upper bound for the true
ground state energy.
\end{itemize}

The simulation results are presented in Fig.\ref{fig:MC}. We observe that the
results predict that $E_0(\la)$ starts deviating significantly form 0 around
$\la\simeq4.6-4.7$, in agreement with the exact enumeration. The very similar
values that we obtain for the ground state energies as obtained by the
simulations for both $N_v=10^3$ and $N_v=10^4$, indicate that the finite size
effects for these sizes of systems, if noticeable, fall well within limitations
of the achievable numerical precision due to the linear cooling scheme. The
results show that $E_0(\la)$ as predicted by the RS approximation is no longer
in agreement with the numerical evidence, thus giving an indirect indication
that one may have to consider a more complicated ansatz for the replica
symmetries. A similar underestimation of the ground state energy in the RS
approximation has been observed in the $K$-SAT problem~\cite{SAT}. In that
model, however, the inconsistency of the RS result was signaled by an
(unphysical) negative ground state energy. This problem was partially solved by
considering a more complicated ansatz for the replica symmetry (i.e. a 1 step
replica symmetry breaking ansatz - 1RSB ). It is therefore plausible that such a
1RSB calculation would also improve on the prediction of the value $\la_c$ at
which the ground state energy ceases to be 0 (i.e. move it closer to the true
value $\la_c\simeq4.7$). Such a calculation (and also subsequent steps in
Parisi's scheme for RSB) is easy to formulate, but its evaluation is numerically
rather involved. This analysis is beyond the scope of the current paper, but
will be the subject of a forthcoming study.
\begin{figure}[h]
\setlength{\unitlength}{1mm}
\begin{picture}(160,50)
\put(  0 , 0 ){\epsfysize=55\unitlength\epsfbox{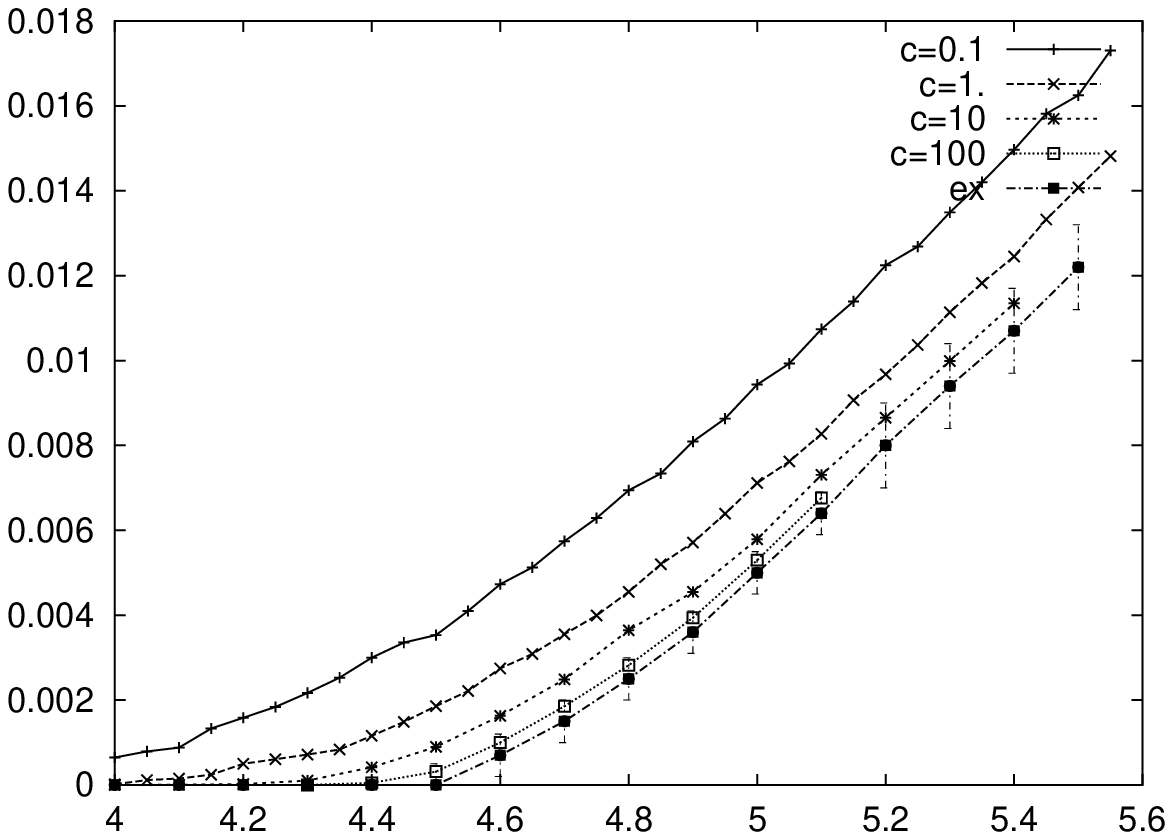}}
\put( 15 ,45 ){\mbox{\fns   \boldmath $E_0(\la)$}}
\put( 30 ,45 ){\mbox{\fns   \boldmath $N_v=10^3$}}
\put( 38 , 0 ){\mbox{\fns   \boldmath $\la$}}
\put( 80 , 0 ){\epsfysize=55\unitlength\epsfbox{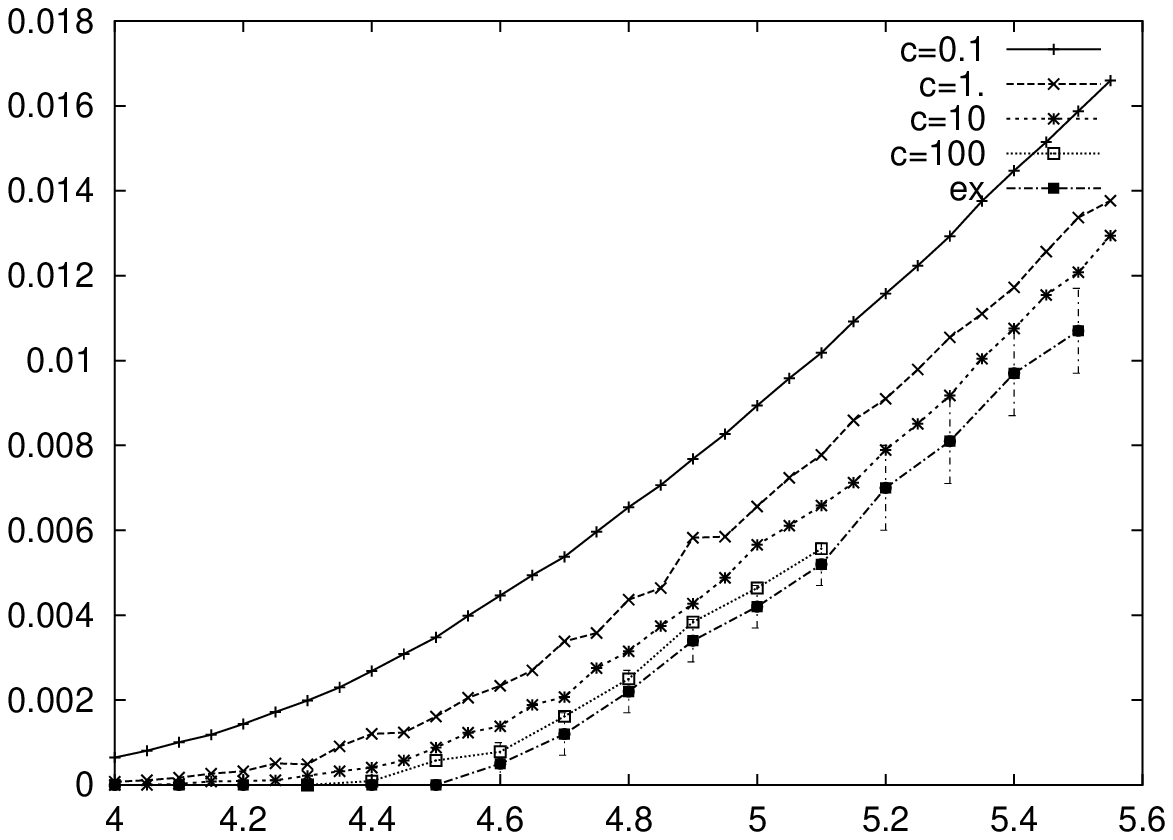}}
\put( 95 ,45 ){\mbox{\fns   \boldmath $E_0(\la)$}}
\put(110 ,45 ){\mbox{\fns   \boldmath $N_v=10^4$}}
\put(115 , 0 ){\mbox{\fns   \boldmath $\la$}}
\end{picture}
\caption{ The ground state energy $E_0(\la)$ as obtained with
MC-simulations with simulated annealing for $N_v=10^3$ (left) and
$N_v=10^4$ (right), and different cooling rates, averaged of 100
runs. The lower curve is the estimate for infinitely slow cooling as
obtained by a quadratic extrapolation of the values for the three
smallest values of $1/C$, to $1/C=0$.
\label{fig:MC}
}
\end{figure}
%
%
\section{2 color problem: exact analysis \label{sec:2E}}
We will now derive an exact expression for the 2-colorability of random graphs,
in the infinite graph size limit, for $\la\in[0,1]$. As we have seen, the
replica analysis correctly finds $E_0(\la)=0$, but is unable to predict the
non-trivial behavior of $\cP_c(\la)$ as observed in the exact enumerations.
We do this by identifying local configurations that give rise to non-colorable
clusters, and by calculating the probabilities of their occurrence. One should
notice that the non-colorable local configurations are loops of odd length:
\begin{figure}[h]
\setlength{\unitlength}{1mm}
\begin{picture}(160,50)
\put( 0  , 5  ){\epsfysize=40\unitlength\epsfbox{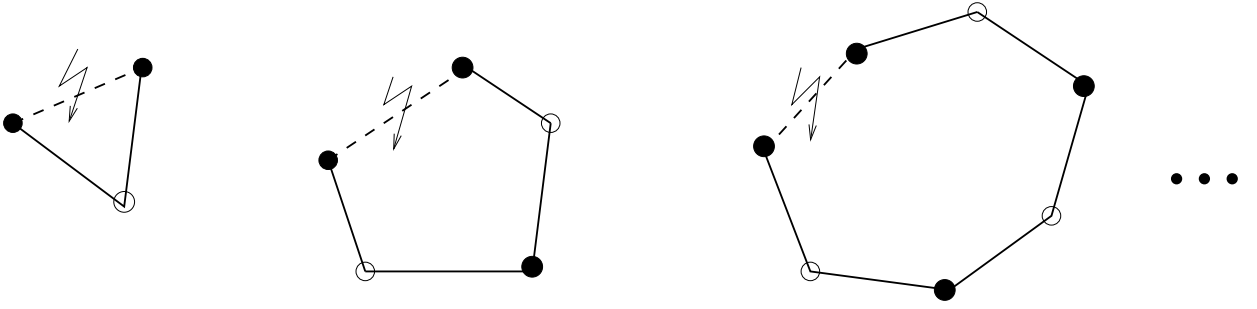}}
\end{picture}
\caption{Loops of odd lengths: 3,5,7,...  all of which have a finite
probability of occurring in large randomly generated graphs for any
finite $\la$.}
\end{figure}
We start from the probability distribution for the number of edges of
a given vertex: $P(L),~L=0,..,\infty$, which is a Poisson
distribution.  We recall from (\ref{D}) that the probabilities of a/no
2-edge between two given vertices are given by
\beq
\cP_e={\la\ov N_{pe/v}}={\la\ov N_v},\hsc 
\cP_{ne}=1-\cP_e\simeq1-{\la\ov N_v} \ .
\eeq
The probability of no (denoted by the symbol $\no$) odd loops in the
graph is given by
\beq
\cP(\no3,\no5,\no7,\no9,..)=\cP(\no3)\cP(\no5|\no3)\cP(\no7|\no3,\no5)
\cP(\no9|\no3,\no5,\no7)..  \ .
\eeq
We first evaluate the probability, that 3 randomly chosen vertices form a loop
of length 3. We randomly pick 3 vertices which can be done in  $\vtd{{N_v}}{3}$
ways. The probability that for a given set of 3 vertices each is connected with
the other two is given by:
\beq
\cP(3)=\cP_e^3={\la^3\ov {N_v}^3}
\eeq
As long as the typical loop size is finite (compared to ${N_v}$), the
correlations between the different $(2k\e1)$-tuples is $\cO({N_v}^{-2})$
(at least 2 new edges have to be present), and are therefore negligible.
Hence, the probability that there are no 3-loops in the graph is given by
\beq
\cP(\no 3)=(1-\cP(3))^{\ftd{{N_v}}{3}}\simeq\exp(-{\la^3\ov 2~3}) \ .
\eeq
Now we turn to the probability that there are no 5-loops, given that there are
no 3-loops. We can randomly pick 5 vertices in $\vtd{{N_v}}{5}$ ways.
The probability that a given set of 5 vertices forms a loop (counting all the
distinct possible orderings ${4!/2}$), while there are no shorter (3-) loops
in the group ($5$ internal edges have to be excluded), is given by:
\beq
\cP(5|\no 3)={4!\ov2}\cP_e^5~\cP_{ne}^5\simeq{4!\ov2}\cP_e^5=\cP(5) \ .
\eeq
Therefore, the probability that there are no 5-loops in the graph is given by
\beq
\cP(\no 5)=(1-\cP(5))^{\ftd{{N_v}}{5}}\simeq\exp(-{\la^5\ov 2~5}) \ .
\eeq
We can repeat this procedure for any odd loop length $2k\e1$, $k=1,2,3..$.
The number of internal edges to exclude is given by $(2k\e1)(2k\m2)/2$, while
the number of distinct orderings of the vertices in a closed loop is given by 
${(2k\e1)!/[2(2k\e1)]}$. Hence, we obtain:
\beq
\cP(\no 2k+1|\no3,..,\no2k-1)\simeq\cP(\no2k+1)\simeq
\exp(-{\la^{2k+1}\ov 2~(2k+1)}) \ .
\eeq 
The probability of no odd loops of any length
(i.e. the probability that the graph is colorable) is therefore:
\beq
\cP_c\simeq \prod_{k=1}^\infty \cP(\no 2k+1)\simeq
\exp\lh-\ha\sum_{k=1}^\infty{\la^{2k\e1}\ov 2k\e1}\rh=
\exp\lh-\ha({\rm atanh}(\la)-\la)\rh=
\lh{1\m\la\ov1\e\la}\rh^{1\ov4}\exp({\la\ov2}) \ .
\eeq 
>From Fig.\ref{fig:eep}, we see that this result is in perfect
agreement with that obtained by exact enumeration up to the point
where the typical odd loop length becomes of the order of the square
root the system size. This point moves to the right (and approaches
$\la=1$) with increasing system size.  Furthermore, since for
$0\leq\la<1$ the probability to have an odd loop is $\cP({\rm
odd})<1$, the ground state energy $E_0$ per edge is then typically
$0$, as the probability to have a finite $E_0$ is exponentially small
in $N_v$:
\beq
P(\cU=E_0>0)\simeq\lh\cP({\rm odd}))\rh^{N_v E_0}\sim 0 \ .
\eeq
This observation is in perfect agreement with our results from the
Monte-Carlo simulations (see Fig.\ref{fig:MC}), and are also confirmed
by our replica analysis.  

Unfortunately, for $p\geq3$ the basic local configurations (i.e. those
including a finite number of vertices) that lead to non-colorability
cannot be enumerated so easily. Furthermore, each of the basic
non-colorable local configurations has a vanishingly small occurrence
probability. It is their collective probability (including very large
configurations that may consist of a finite fraction of the graph)
that suddenly becomes 1 at the critical $\la$, giving rise to the
observed 1st order transition from colorable to non-colorable graphs.
\begin{figure}[h]
\setlength{\unitlength}{1mm}
\begin{picture}(160,50)
\put( 0  , 5  ){\epsfysize=40\unitlength\epsfbox{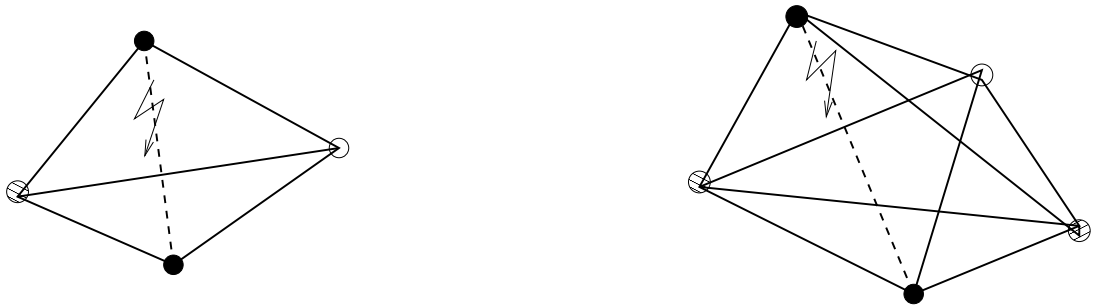}}
\end{picture}
\caption{The smallest elementary non-colorable configurations for
$p=3$ (left), $p=4$ (right), both of which have a vanishingly small
occurrence probability in large randomly generated graphs for any
finite $\la$.}
\end{figure}
%

%

\section{Conclusions \label{sec:CO}}
We analyzed the colorability of random graphs for finite average connectivity,
an important NP-complete problem.  The statistical physics based analysis
provides typical results in the infinite system size limit, complementing
results published in the computational complexity literature.

The results obtained are in qualitative agreement with those reported
in the literature as well as with numerical results we obtained from exact
enumeration and Monte-Carlo based solutions. 

One apparent discrepancy, in the case of 2-color graphs, has been investigated
using a probabilistic analysis that provided exact results for the probability
of colorable random graphs in the case of two colors. The analysis also explains
the failure of the statistical physics based analysis to detect uncolorability
when this comes as a result of only a finite number of unsatisfiable edges,
since such an analysis can identify only an extensive number of such edges.

The current analysis is the first step in the investigation of graph
colorability. Future studies include: a)~Refining the current analysis
by extending it to the case of 1-step RSB. b)~Investigating graphs
with mixed 2 and 3-color vertices; this case has been studied
numerically in~\cite{Walsh} but is difficult to analyze due to the
different nature of two analyzes. c)~Studying the properties of random
graphs with various restrictions. These research activities are
currently underway.

\vspace{3mm}
\noindent
Acknowledgments: \nl {\footnotesize We would like to thank Toby Walsh
for introducing us to the problem and for useful discussions.  Support by
The Royal Society and EPSRC-GR/N00562 is acknowledged.  }


\appendix
\section{technical details of the replica calculation \label{A}}
%
We now present the technical details of the replica calculation. We
calculate the average of the $\rn$-th power of the partition sum:
\beq
\cZ^n(\vf,\cD)\ev\prod_{a=1}^n\lv
\Tr_{\vcc^a}\exp\lh-\be\sum_\BKK\cD_\BKK~\chi^a_\BKK(\vcc)\rh
\prod_\mu\de(\sum_{j=1}^N \mu_{c^a_j}-Nf_\mu)\rv \ .
\label{A:Z}
\eeq
The constraints on the $f_\mu$ are enforced by the introduction of the
Lagrange multipliers $\htf^a_\mu$, such that the average of the
replicated partition sum becomes
\bea
\Bra\cZ^n \Ket&=&
\int\prod_{a=1}^n\prod_{\mu=1}^p
\lc{d\htf^a_\mu\ov 2\pi\ri}\exp\lh -Nf_\mu\htf^a_\mu\rh\rc
\prod_{a=1}^n\lc\Tr_{\vcc^a}\exp\lh\sum_{\mu=1}^p\sum_{j=1}^{N_v}\htf^a_\mu
\mu_{c^a_j}\rh\rc\times\nn \\&&\hsc
\Bra\prod_{a=1}^n\exp\lh-\be\sum_\BKK\cD_\BKK~\chi_\BKK(\vcc^a)\rh\Ket_\cD \ .
\label{A:Zn}
\eea
The average over all tensors $\cD$ with $K$ (taken to be 2 for now) non zero 
elements per row, and $L_j$ per column $j$ is given by (\ref{avD}), where 
the Kronecker deltas can be expressed as $\de(x,y)=\oint{dZ\ov2\pi\ri}
Z^{(x-y-1)}$. We now proceed with the calculation of $\cT$
\bea
\cT&=&\Tr_\cD\oint\prod^{N_v}_{j_1=1}\lc{dZ_{j_1}\ov2\pi\ri }
Z_{j_1}^{(\sum_{\Bra j_2,..,j_K\Ket}\cD_\BKK-L_{j_1}-1)}\rc
\prod_\BKK\prod_{a=1}^\rn\exp\lh-\be\cD_\BKK~\chi_\BKK(\vcc^a))\rh         \nn\\
&=&\oint\prod^{N_v}_{j=1}\lc{dZ_j\ov2\pi\ri }Z_j^{-(L_j\e 1)}\rc
\prod_\BKK\lc\Tr_{\cD_\BKK}[Z]_\BKK^{\cD_\BKK}
\prod_{a=1}^\rn\exp\lh-\be\cD_\BKK~\chi_\BKK(\vcc^a)\rh\rc                 \nn\\
&=&\oint\prod_{j=1}^{N_v}\lc{dZ_j\ov2\pi\ri }Z_j^{-(L_j\e 1)}\rc
\exp\lh\sum_\BKK\log\lv1+[Z]_\BKK\prod_{a=1}^\rn
\exp\lh-\be~\chi_\BKK(\vcc^a)\rh\rv\rh                                     \nn\\
&\simeq&\oint\prod_{j=1}^{N_v}\lc{dZ_j\ov2\pi\ri }Z_j^{-(L_j\e 1)}\rc 
\exp\lh\sum_\BKK[Z]_\BKK\prod_{a=1}^\rn\exp\lh-\be~\chi_\BKK(\vcc^a)\rh\rh
\label{step1}                                                                 \\
&=&\oint\prod_{j=1}^N\lc{dZ_j\ov2\pi\ri }Z_j^{-(L_j\e 1)}\rc 
      \exp\lh\sum_\BKK[Z]_\BKK\prod_{a=1}^\rn
      \lv1-\sum_{\mu=1}^p\De~[\mu_{c_j^a}]_\BKK\rv\rh                          
\label{step2}                                                                 \\
&=&\oint\prod_{j=1}^N\lc{dZ_j\ov2\pi\ri }Z_j^{-(L_j\e 1)}\rc
   \exp\lh\sum_{\rmm=0}^\rn(-\De)^\rmm\sum_\BKm\sum_\BKmu\sum_\BKK\lv
   Z[\mu_{c_j}]^\BKm_\BKmu\rv_\BKK\rh                                      \nn
\eea
where we have used the short hand notations (\ref{BKm}).  Step
(\ref{step1}) is justified, because after integration over the $Z_j$,
only those terms in the expansion of the exponential in which each
$Z_j$ occurs exactly $L$ times will survive, and it was
shown~\cite{Wong} that in the thermodynamic limit ($N_v\to\infty$), in
the expansion of the logarithm all higher order terms are negligible
compared to the first order term. In step (\ref{step2}), we have made
the choice (\ref{Xi}) for $\chi^a_\BKK$.  \nl We have that
$\sum_\BKK[x]_\BKK\simeq\lh\sum_{j=1}^{N_v}x_j\rh^K/K!$~, in the
thermodynamic limit, such that
\beq
\cT\simeq\oint\prod_{j=1}^{N_v}\lc{dZ_j\ov2\pi\ri }Z_j^{-(L_j\e 1)}\rc \exp\lh
{1\ov K!}\sum_{\rmm=0}^n(-\De)^\rmm\sum_\BKm\sum_\BKmu
\lv\sum_{j=0}^{N_v}Z_j[\mu_{c_j}]^\BKm_\BKmu\rv^K\rh \ .
\eeq
In order to factorize the whole expression in the $j$'s, we introduce the
order parameters 
\beq
q^\BKm_\BKmu\ev\sum_{j=0}^NZ_j~[\mu_{c_j}]^\BKm_\BKmu,
\label{A:qdef}
\eeq
by the introduction of the corresponding Lagrange multipliers
$\htq^\BKm_\BKmu$.
\beq
\cT=\int\prod_{\rmm=0}^\rn\prod_\BKm\prod_\BKmu
\lc{d\htq^\BKm_\BKmu dq^\BKm_\BKmu\ov2\pi\ri}
\exp\lh-\htq^\BKm_\BKmu q^\BKm_\BKmu+(-\De)^\rmm{(q^\BKm_\BKmu)^K\ov K!}\rh\rc
\prod_{j=1}^{N_v}X_j,
\eeq
where
\beq
X_j=\oint{dZ_j\ov2\pi\ri}Z_j^{-(L_j\e 1)}\exp\lh Z_j
\sum_{m=0}^n\sum_\BKm\sum_\BKmu
\htq^\BKm_\BKmu[\mu_{c_j}]^\BKm_\BKmu\rh
={1\ov L_j!}\lh\sum_{m=0}^n\sum_\BKm\sum_\BKmu
\htq^\BKm_\BKmu~[\mu_{c_j}]^\BKm_\BKmu\rh^{L_j}
\eeq
Following similar steps we obtain for the denominator
\beq
\cN\simeq\int\lc{d\htq_0dq_0\ov2\pi\ri}\rc\exp\lh-\htq_0q_0+{q_0^K\ov K!}+
N\sum_LP(L)\log\lh{\htq_0^L\ov L!}\rh\rh
\label{A:N}
\eeq
The average of the replicated partition function hence reads
\bea
\Bra\cZ^\rn\Ket&=&{1\ov\cN}
\int\prod_{a=1}^\rn\prod_{\mu=1}^p\lc{d\htf_\mu^a\ov2\pi\ri}
                     \exp\lh-N\htf^a_\mu f_\mu\rh\rc\nn\\
&&\hsc\prod_{\rmm=0}^\rn\prod_\BKm\prod_\BKmu
 \lc{d\htq^\BKm_\BKmu dq^\BKm_\BKmu\ov2\pi\ri}\exp\lh
 -\htq^\BKm_\BKmu q^\BKm_\BKmu+(-\De)^\rmm{(q^\BKm_\BKmu)^K\ov K!}\rh\rc\nn\\
&&\hsc\prod_{j=1}^{N_v}\prod_{a=1}^\rn~\Tr_{c_j^a}~
\exp(\sum_\mu\htf^a_\mu\mu_{c^a_j})\lc
{1\ov L_j!}\lh \sum_{\rmm=0}^\rn\sum_\BKm\sum_\BKmu
             \htq^\BKm_\BKmu[\mu_{c_j}]^\BKm_\BKmu \rh^{L_j}\rc
\label{A:Zng}
\eea
which can be evaluated using the saddle point method for the integration
variables $\htf^a_\mu$, $\htq^\BKm_\BKmu$ and $q^\BKm_\BKmu$. In order to
proceed with the calculation, we must make an assumption about the symmetry
between replicas, and we use the replica symmetric ansatz (\ref{RS}) for the
terms in (\ref{A:Zng}) that involve the order parameters:
\bea
\sum_{\rmm=0}^\rn\sum_\BKm\sum_\BKmu\htq^\BKm_\BKmu q^\BKm_\BKmu&=&
 q_0\htq_0\int'\{d\vx d\vhx~\pi(\vx)~\htpi(\vhx)\}
 \sum_{\rmm=0}^\rn\sum_\BKm\sum_\BKmu\prod_\mu(-x_\mu\htx_\mu)^{m_\mu}     \nn\\
&=&
 q_0\htq_0\int'\{d\vx d\vhx~\pi(\vx)~\htpi(\vhx)\}\sum_{\rmm=0}^\rn
 \vtd{\rn}{\rmm}\sum_\vm\vtd{\rmm}{\vm}\prod_\mu(-x_\mu\htx_\mu)^{m_\mu}   \nn\\
&=&
 q_0\htq_0\int'\{d\vx d\vhx~\pi(\vx)~\htpi(\vhx)\}
 \sum_{\rmm=0}^\rn\vtd{\rn}{\rmm}(-\sum_\mu x_\mu\htx_\mu)^\rmm            \nn\\
&=&
q_0\htq_0\int'\{d\vx d\vhx~\pi(\vx)\htpi(\vhx)\}~(1-\sum_\mu x_\mu\htx_\mu)^\rn\\
\sum_{\rmm=0}^\rn(-\De)^\rmm\sum_\BKm\sum_\BKmu{(q^\BKm_\BKmu)^K\ov K!}&=&..=
  {q_0^K\ov K!}\int'\prod_{k=1}^K\{d\vx_k~\pi_k(\vx_k)\}
  (1-\De\sum_\mu\prod_{k=1}^Kx_{k,\mu})^\rn                                   \\
\sum_{\rmm=0}^\rn\sum_\BKm\sum_\BKmu\htq^\BKm_\BKmu[\mu_{c_j}]^\BKm_\BKmu&=&
 \htq_0\int'\{d\vhx~\htpi(\vhx)\}\sum_{\rmm=0}^\rn\sum_\BKm\sum_\BKmu
 \prod_{\mu=1}^p(-\htx_\mu)^{m_\mu}\prod_{\ell=1}^m\mu_{\ell~c^{a_\ell}_j} \nn\\
&=&
 \htq_0\int'\{d\vhx~\htpi(\vhx)\}\sum_{\rmm=0}^\rn\sum_\BKm\lh
 \sum_\BKmu\prod_{\ell=1}^\rmm(-\mu_{\ell~c^{a_\ell}_j}~\htx_{\mu_\ell})\rh\nn\\
&=&
 \htq_0\int'\{d\vhx~\htpi(\vhx)\}\sum_{\rmm=0}^\rn\sum_\BKm
 \lh\prod_{\ell=1}^\rmm(-\sum_\mu\mu_{\ell~c^{a_\ell}_j}~\htx_\mu)\rh      \nn\\
&=&..=
\htq_0\int'\{d\vhx~\htpi(\vhx)\}\prod_{a=1}^\rn(1-\sum_\mu\mu_{c_j^a}~\htx_\mu)~,
\eea
where $\vtd{\rmm}{\vm}~(\ev{\rmm!/\prod_\mu m_\mu!})$~ are multi($p$)-nomial-~,
and $\vtd{\rn}{\rmm}~(\ev{\rn!/\rmm!(\rn-\rmm)!})$ binomial coefficients.
Hence, we have 
\bea
\prod_{a=1}^\rn\lc\Tr_{c_j^a}~\exp(\sum_\mu\htf_\mu\mu_{c^a})\rc{1\ov L_j!}
\lh\cdots\rh^{L_j}&=&                                                      \nn\\
&&\hspace{-50mm}={\htq_0^{L_j}\ov L_j!}
\int'\prod_{l=1}^{L_j}\{d\vhx_l~\htpi_l(\vhx_{l,\mu})\}\prod_{a=1}^\rn\lh
\Tr_{c_j^a}\exp(\sum_{\mu=1}^p\htf_\mu\mu_{c^a})
\prod_{l=1}^{L_j}(1-\sum_\mu \mu_{c^a}~\htx_{l,\mu})\rh                    \nn\\
&&\hspace{-50mm}={\htq_0^{L_j}\ov L_j!}
\int'\prod_{l=1}^{L_j}\{d\vhx_l~\htpi_l(\vhx_{l,\mu})\}\lh
\sum_{\mu=1}^p~\exp(\htf_\mu)\prod_{l=1}^{L_j}(1-\htx_{l,\mu})\rh^\rn \ ,
\eea
to obtain the following expression for the averaged replicated partition sum
\beq 
\Bra\cZ^\rn\Ket={1\ov\cN }\ext_{\{\vhf,\htq,q,\htpi,\pi\}}\exp\lv
-\rn N_v\sum_{\mu=1}^pf_\mu\htf_\mu-q_0\htq_0~\cI_1+{q_0^K\ov K!}\cI_2+N_v\sum_LP(L)\lh\log({\htq_0^L\ov L!})+\log(\cI_{3L})\rh\rv ~,
\label{A:ZnRS}
\eeq
where
\bea
\cI_1&\ev&\int'\{d\vx d\vhx~\pi(\vx)~\htpi(\vhx)\}~
(1-\sum_\mu x_\mu\htx_\mu)^\rn                                             \nn\\
\cI_2&\ev&\int'\prod_{k=1}^K\{d\vx_k~\pi_k(\vx_k)\}~
(1-\De\sum_\mu \prod_{k=1}^Kx_{k,\mu})^\rn                         \label{IRS}\\
\cI_{3L}&\ev&\int'\prod_{l=1}^L\{d\vhx_l~\htpi_l(\vhx_l)\}
\lh\sum_{\mu=1}^p~\exp(\htf_\mu)\prod_{l=1}^L(1-\htx_{l,\mu})\rh^n         \nn \ .
\eea
We now solve the saddle point equations with respect to $\htq_0$ and
$q_0$, and note that the structure of the $(\htq_0,q_0)$-dependent
part of the denominator is exactly the same with $\cI_1=\cI_2=1$, to
obtain
\beq
\lc\bay{l}      q_0=\lh{N_v\la(K-1)!\ov\cI_2}\rh^{1/K}\\
             \htq_0={N_v\la \ov\cI_1}\lh{\cI_2\ov N_v\la (K-1)!}\rh^{1/K}\eay\rp
\hsc\to\hsc
\lc\bay{l}q_0\htq_0={N_v\la \ov\cI_1}\\
      {q_0^K\ov K!}={N_v\la \ov K\cI _2}\eay\rp  \ ,
\eeq
where $\la=\sum_LP(L)~L$, such that all terms not depending on the
$\cI_i$ or $f_\mu$ in the numerator and denominator cancel:
\beq 
\Bra\cZ^\rn\Ket\simeq\exp\lv N_v\lh-\rn\sum_{\mu=1}^pf_\mu\htf_\mu
-\la \log(\cI_1)+{\la \ov K}\log(\cI_2)+\sum_LP(L)\log(\cI_{3L})\rh\rv  \ ,
\eeq 
taken in the extremum for $\{\vhf,\htpi,\pi\}$. So far we have
performed all calculations for general positive integer $\rn$. Taking
$\lim_{\rn\to0}{ (\cZ^\rn-1)\ov\rn}$, and multiplying the result with
${K\ov N_v\la}$, we obtain the replica symmetric free energy per edge
(\ref{FRS}).
%
\section{low temperature limit \label{B}}
We will now show that even in the limit $\be\to\infty$, the distribution
$\pi(\vx)$ remains non-trivial. In order to demonstrate this, we concentrate on
the fixed point equation (\ref{dhpi2}). Using two explicit examples, we show how
contributions to $\pi()$ for extremal values of the arguments (i.e.
$1-x_\mu\ev\eps_\mu=\cO(\exp(-\be)$,~$x_\nu=\cO(\exp(-\be))~|~\nu\neq\mu$)
may generate contributions to $\pi()$ with finite argument values (i.e.
$1-x_\mu=\cO(1),~\forall~\mu$) and vice-versa.\nl 

$\bullet~1)$ First, we assume that there is a finite probability density
$\pi(\vx)$ that $1-x_\mu\ev\eps_\mu=\cO(\exp(-\be))$, such that
$x_\nu=\cO(\exp(-\be))~|~\nu\neq\mu$~. Suppose now that $p=3$, and consider the
term in (\ref{dhpi2}) with $L\m1=3$. The following combination of $\vx_\ell$'s
($\ell=1,2,3$) has then a finite probability density:
\beq
\lc
\bay{ll}
1-x_{1,1}\ev\eps_{1,1},~~1-x_{2,2}\ev\eps_{2,2},~~1-x_{3,3}\ev\eps_{3,3},\hsc &
\eps_{i,i}=\cO(\exp(-\be))                                                \\ \\ 
x_{\ell,\nu}=\cO(\exp(-\be))\hsc &\nu\neq\mu
\eay
\rp
\eeq
and, to leading order, generates a contribution to the l.h.s. of (\ref{dhpi2})
with $\vx$:
\beq
1-x_\mu\simeq1-{\exp(\m\be)+\eps_{\mu,\mu}\ov
               \sum_\nu(\exp(\m\be)+\eps_{\nu,\nu})}=\cO(1)\ ,\hsc \mu=1,2,3\ ,
\eeq 
i.e. with finite argument values.

$\bullet~2)$ Second, we assume that there is a finite probability density 
$\pi(\vx)$ that $1-x_\mu\ev\eps\ll 1$, such that
$x_\nu=\cO(\eps)~|~\nu\neq\mu$~. Suppose now that $p=2$, and consider the term
in (\ref{dhpi2}) with $L\m1=3$. The following combination of $\vx_\ell$'s
($\ell=1,2,3$) has then a finite probability density:
\beq
\lc
\bay{ll}
1-x_{1,1}\ev\eps_{1,1},~~1-x_{2,1}\ev\eps_{2,1},\hsc&\eps_{1/2,1}=\cO(\eps)\\\\ 
x_{3,\nu}=\cO(1)~~&\nu=1,2
\eay
\rp
\eeq
and, to leading order, generates a contribution to the l.h.s. of (\ref{dhpi2})
with $\vx$: 
\beq
x_1\simeq{\eps_{1,1}\eps_{2,1}\ov\eps_{1,1}\eps_{2,1}+1\m x_{3,2}}=\cO(\eps^2),
\hsc
1-x_2\simeq1-{1\m x_{3,2}\ov\eps_{1,1}\eps_{2,1}+1\m x_{3,2}}=\cO(\eps^2).
\eeq 
i.e. with more extreme values of the arguments (~$\cO(\eps^2)$ instead of 
$\cO(\eps)$~).
\nl\nl
Hence, we have shown that extreme values will generate less extreme values
and vice-versa. Since the r.h.s of (\ref{dhpi2}) contains terms with all values
of $L$, obviously (even in the limit $\be\to\infty$) we cannot explicitly keep
track of the proliferation of distributions to different values of $\vx$, and
have to resort to a numerical analysis. For each value of $\la$, we have to
check whether in the limit $\be\to\infty$ a finite probability density
$\pi(\vx)$ is generated for extremal values of $\vx$ (i.e. $1-x_\mu=
\cO(\exp(-\be))$). If this is the case, the internal energy $\cU$ will be
positive, and the probability that the graph is colorable must be 0.
\end{document}